\documentclass[screen,prologue,dvipsnames,acmsmall]{acmart}
\AtBeginDocument{%
  \providecommand\BibTeX{{%
    \normalfont B\kern-0.5em{\scshape i\kern-0.25em b}\kern-0.8em\TeX}}}

\setcopyright{acmlicensed}
\acmJournal{PACMHCI}
\acmYear{2025} \acmVolume{1} \acmNumber{CSCW}
\acmDOI{XXXXXXX.XXXXXXX}

\usepackage{xcolor}
\usepackage{multirow}
\usepackage[flushleft]{threeparttable}
\usepackage{caption}
\usepackage{subcaption}
\usepackage{tabularx}
\usepackage[most]{tcolorbox}

\begin{document}

\title[Privacy and Security Micro-Lessons]{Creating and Evaluating Privacy and Security Micro-Lessons for Elementary School Children}

\author{Lan Gao}
\email{langao@uchicago.edu}
\affiliation{%
  \department{Department of Computer Science}
  \institution{University of Chicago}
  \city{Chicago}
  \state{Illinois}
  \country{USA}
}

\author{Elana B Blinder}
\email{eblinder@umd.edu}
\affiliation{%
\department{College of Information Studies}
  \institution{University of Maryland}
  \city{College Park}
  \state{Maryland}
  \country{USA}}

\author{Abigail Barnes}
\email{abigailbarnes@uchicago.edu}
\affiliation{%
  \department{Department of Computer Science}
  \institution{University of Chicago}
  \city{Chicago}
  \state{Illinois}
  \country{USA}
}

\author{Kevin Song}
\email{ksong814@uchicago.edu}
\affiliation{%
  \department{Department of Computer Science}
  \institution{University of Chicago}
  \city{Chicago}
  \state{Illinois}
  \country{USA}
}

\author{Tamara Clegg}
\email{tclegg@umd.edu}
\affiliation{%
\department{College of Information Studies / College of Education}
  \institution{University of Maryland}
  \city{College Park}
  \state{Maryland}
  \country{USA}}

\author{Jessica Vitak}
\email{jvitak@umd.edu}
\affiliation{%
\department{College of Information Studies}
  \institution{University of Maryland}
  \city{College Park}
  \state{Maryland}
  \country{USA}}

\author{Marshini Chetty}
\email{marshini@uchicago.edu}
\affiliation{%
  \department{Department of Computer Science}
  \institution{University of Chicago}
  \city{Chicago}
  \state{Illinois}
  \country{USA}
}

\renewcommand{\shortauthors}{Gao et al.}

\newcommand{\todo}[1]{\textcolor{purple}{LG: #1}}
\newcommand{\mc}[1]{\textcolor{orange}{MC: #1}}
\newcommand{\tc}[1]{\textcolor{red}{TLC: #1}}
\newcommand{\tlc}[1]{\textcolor{red}{TLC: #1}}
\newcommand{\jv}[1]{\textcolor{green}{JV: #1}}
\newcommand{\eb}[1]{\textcolor{green}{EB: #1}}

\newcommand{\implementer}[1]{\tcbox[on line, 
        boxsep=0pt, boxrule=0pt, left=2pt,right=2pt,top=0pt,bottom=-1pt,
        colframe=white,colback=Aquamarine!25!, fontupper={\strut}]{#1}%
        }
\newcommand{\commenter}[1]{\tcbox[on line, 
        boxsep=0pt, boxrule=0pt, left=2pt,right=2pt,top=0pt,bottom=-1pt,
        colframe=white,colback=Salmon!35!, fontupper={\strut}]{#1}%
        }
\newcommand{\formative}[1]{\tcbox[on line, 
        boxsep=0pt, boxrule=0pt, left=2pt,right=2pt,top=0pt,bottom=-1pt,
        colframe=white,colback=Orchid!27!, fontupper={\strut}]{#1}%
        }

\begin{abstract}
  The growing use of technology in K--8 classrooms highlights a parallel need for formal learning opportunities aimed at helping children use technology safely and protect their personal information. Even the youngest students are now using tablets, laptops, and apps to support their learning; however, there are limited curricular materials available for elementary and middle school children on digital privacy and security topics. To bridge this gap, we developed a series of micro-lessons to help K--8 children learn about digital privacy and security at school. We first conducted a formative study by interviewing elementary school teachers to identify the design needs for digital privacy and security lessons. We then developed micro-lessons---multiple 15-20 minute activities designed to be easily inserted into the existing curriculum---using a co-design approach with multiple rounds of developing and revising the micro-lessons in collaboration with teachers. Throughout the process, we conducted evaluation sessions where teachers implemented or reviewed the micro-lessons. Our study identifies strengths, challenges, and teachers' tailoring strategies when incorporating micro-lessons for K--8 digital privacy and security topics, providing design implications for facilitating learning about these topics in school classrooms. 
\end{abstract}





\maketitle

\section{Introduction}

Children are navigating digital spaces at younger and younger ages \cite{Auxier_2020, Erikson_2022}. While much of the focus on this technology use has been its potential educational benefits, as well as concerns about too much screen time, it also highlights a critical need for children to begin learning digital privacy and security concepts---specifically, the ability to identify, evaluate, and respond to privacy and security risks children may face in digital spaces, as well as strategies that help them use technology and behave in ways that maintain personal privacy, security, and safety. 
Over the last decade, child-computer interaction (CCI) and learning sciences researchers have explored how children could deepen their understanding of digital privacy and security through informal learning approaches, such as interacting with privacy education games and e-books~\cite{raynes2014gaming,maqsood2021design,kumar2018co, zhang2017cyberheroes,yap2020phone}. These studies highlight advances in informal learning approaches designed to strengthen children's awareness of online risks, such as cyberbullying and information over-disclosure, and promoting protective actions against these risks. 


While these studies showcase the utility of informal learning opportunities, we argue there is significant potential for more formalized learning about digital privacy and security in elementary and middle school (Grades K--8) classrooms. 
First, compared to extracurricular activities and after-school programs, children spend most of their time learning in the classroom, which opens up a large opportunity for schools to become central to helping children develop digital privacy and security literacy~\cite{lastdrager2017effective, rahman2020importance}. Second, researchers have suggested that children can learn digital privacy and security concepts through socialization in group activities and with peer support~\cite{nicholson2021understanding}, which could be best supported through social dynamics at school. Moreover, with the increasing adoption of educational technology in school learning, CSCW researchers have studied digital privacy and security practices and tensions at school, such as privacy and security challenges of technology use in classroom~\cite{kumar2019privacy,chanenson2023uncovering,cino2020does,bacak2022elementary,lu2021coding} and the mitigation of risks arising from these challenges~\cite{maqsood2021they, martin2023teacher}, as well as privacy and security conflicts among students, parents, and teachers in remote learning~\cite{wagman2023we}. 

When considering the potential gains of in-classroom privacy and security learning and expanding sociotechnical infrastructure at school, it is crucial to incorporate digital privacy and security education into the classroom environment starting in elementary school. Yet there has been little focus in prior works on developing privacy and security lessons, particularly for younger children. And when such lessons exist, they generally focus on a single, narrow topic---for example, AI privacy~\cite{10.1145/3613904.3642460}, digital citizenship~\cite{commonsense_digital_citizenship, li2023integrating}, or critical data literacy~\cite{bilstrup2022supporting}---rather than covering multiple, foundational topics that build upon each other. Moreover, because digital privacy and security are not often part of the formal curriculum,
teachers struggle to find open-source, age-appropriate, and ready-to-use lessons~\cite{lamond2022sok, sauglam2023systematic, kumar2023exploring}, or create new digital lessons areas~\cite{kumar2019privacy}. Finally, traditional training, which usually offers a one-off learning experience, may fail to maintain children's digital privacy and security awareness persistently~\cite{lastdrager2017effective}. 
These challenges highlight the need to develop a digital privacy and security curriculum oriented toward K--8 that is easy to adopt, fits within the limited scope of time and space at school, and fosters an enduring understanding of related topics over time.

To address this gap in digital privacy and security education, we took an iterative co-design approach, working with teachers to develop micro-lessons for K--8 educators that help children learn digital privacy and security concepts in the classroom. We sought to answer the following research questions:


\begin{itemize}
    \item RQ1: What do K--8 teachers need to help children learn digital privacy and security concepts? 
    \item RQ2: How can digital privacy and security micro-lessons be integrated into K--8 teachers' existing lesson plans? 
    \item RQ3: From K--8 teachers' perspective, how can micro-lessons help K--8 children engage with digital privacy and security concepts? 
\end{itemize}





To answer these questions, we first 
conducted seven interviews with 14 STEM and non-STEM teachers from two partnering public elementary schools in two US metropolitan areas, to inform the design of the privacy and security lessons. Our findings revealed five design needs for lesson structure and content: 1) including various aspects of digital privacy and security in the lessons; 2) differentiating learning goals and lesson formats to children in different grade bands; 3) integrating everyday relatable contexts in the lessons; 4) constructing short and robust lessons; and 5) providing effective and simple teaching resources for teachers in the lessons (RQ1).  

Next, we iteratively developed a set of micro-lessons via co-design~\cite{severance2018organizing, sanders2008co} and evaluation with K--8 teachers teaching diverse STEM and non-STEM subjects with varied elementary education experience. 
To evaluate our micro-lessons, seven teachers implemented the initial version of micro-lessons, then completed interviews and shared their experiences and feedback. Feedback from these initial evaluations was used to further iterate on the micro-lessons and improve their structure and content. Six additional teachers then provided external reviews of the updated micro-lessons and provided additional feedback on their feasibility in the classroom, yielding three key findings. First, the micro-lesson format was seen as flexible and easy to integrate into the existing curriculum. At the same time, infrastructural issues may mean the resources within the micro-lessons are not available for all schools (RQ2). Second, the micro-lessons provided engaging ways for children to learn about privacy and security. However, teachers suggested they could be improved by tailoring lessons for different student needs and by including more professional development materials for teachers. We also found evidence that both children's and teachers' awareness of digital privacy and security concepts could be improved after engaging with the micro-lessons (RQ3).


This research extends prior CSCW research on sociotechnical systems and learning~\cite{blinder2024evaluating,schwab1992,Slovak2016,sobel2017,wagman2023we}, as well as a long tradition of social computing research addressing social aspects of privacy and security (e.g., \cite{jackson2014,kumar2017no,palen2004, wisniewski2015}). This research also contributes to the Computer Science Education research on privacy and security education (e.g., \cite{culnan2009online,vsvabensky2020cybersecurity}), which has primarily focused on higher education while lacking a perspective on elementary education. Specifically, this study 1) provides empirical evidence regarding the need for and challenges in implementing privacy and security curriculum, particularly with very young children, and 2) presents a micro-lesson curriculum geared toward Grades K--8 and spanning four topics around digital privacy and security.  
Following works by Kumar and colleagues~\cite{kumar2017no,kumar2019privacy,kumar2020strengthening}, the micro-lessons emphasize learning opportunities across school and home contexts through authentic examples that resonate with children’s lived experiences. The lessons move beyond basic ``do's and don'ts'' to include activities that help children discuss and reflect on how privacy and security manifest in their everyday lives. They also allow children to continually interrogate how digital privacy and security fit into their lives over time, as their use of technologies evolves.

\section{Related Work}




In this section, we review prior work investigating children's digital privacy and security literacy, building digital privacy and security education approaches for children, and digital privacy and security education in K--8 schools.

\subsection{Children's Digital Privacy and Security Literacy}
Much of the research on children's digital privacy and security focuses on safety from online threats. Failing to maintain digital privacy and security exposes children to online threats~\cite{nolan2011stranger, livingstone2006children}. 
Threats like cyberbullying are increasing for children of all ages, and younger children are getting exposed to them earlier. Researchers have categorized risks around a `3C' framework (see ~\cite{hasebrink2009comparing}) based on the types of \textit{content} children may access, with whom they \textit{contact}, and their \textit{conduct} when interacting with others online.
Younger children are particularly vulnerable to online threats because they may not fully understand the consequences of interacting with technology. Studies show that children are especially unaware of password security~\cite{lamichhane2017investigating, maqsood2018exploratory}, phishing attacks~\cite{tirumala2016survey, lastdrager2017effective, nicholson2020investigating}, and datafication pipeline~\cite{ge2022dont}. Until recently, however, there has been limited evaluation of elementary-aged children's digital privacy and security literacy \cite{kumar2017no, kumar2020strengthening}. 

Researchers have also outlined the difficulties of helping young children learn about privacy and security. First, although many children have a basic awareness of digital privacy and security, it is difficult for them to understand more complex concepts related to technology 
~\cite{agesilaou2022whose,livingstone2020data}. Children can be easily fooled by certain types of content---Zhao et al.~\cite{zhao2019make} found children have less awareness of the risks of game promotions and advertisements compared to other threats such as requesting sensitive personal information. Lastly, children may struggle to integrate what they learn 
into practice in the long term~\cite{lamond2022sok, lastdrager2017effective}. As a result, children may blindly trust digital interactions and overshare information without sufficient digital privacy and security literacy~\cite{mcreynolds2017toys,yan2021risk}. 

Parents and teachers are largely responsible for children's online safety, especially for young children who may have limited awareness of digital privacy and security~\cite{assal2018exploration, kumar2017no, maqsood2021they, wagman2023we}. Prior work suggests that both parents and teachers take steps to help children learn about digital privacy and security~\cite{williams2023youth, kumar2019privacy, liu2024integrating}. At the same time, however, several barriers exist. Parents~\cite{liu2024integrating} and teachers~\cite{kumar2019privacy} believe they lack sufficient training and literacy to educate children about digital privacy and security. Some parents employ a strategy of focusing on controlling online activities,  rather than conversation or education~\cite{williams2023youth, zhang2016nosy, liu2024integrating}, limiting children's self-autonomy. Teachers may lack educational resources like professional development on related topics, and time usually restricts them from conducting in-classroom digital privacy and security education~\cite{kumar2019privacy, maqsood2021they, drader2022digital, smith2023educators, mcleod2024comparing}.

CCI researchers, in particular, have called for actions to enhance children's knowledge and practice in digital privacy and security (e.g., \cite{kumar2020strengthening}). 
To date, however, there has been limited work addressing the learning and teaching needs necessary to improve children's literacy, including the topics children should learn and the requirements for developing educational interventions. To bridge this gap, our work explores teachers' curricular needs and develops privacy and security-focused micro-lessons that can be used in K--8 classrooms. 




\subsection{Digital Privacy and Security Education Approaches for Children}
Recent years have seen an increase in the number and diversity of educational tools that address digital privacy and security
~\cite{zhang2021systematic}. For example, Google~\cite{google_be_internet_awesome} and Meta~\cite{meta_youth_safety} have built programs focusing on digital citizenship and online safety. 
Likewise, Common Sense Media~\cite{commonsense_digital_citizenship}, a non-profit focused on children's media and technology, provides numerous educational resources on digital citizenship which are used by over 60,000 schools in the US~\cite{james2019teaching}.

Researchers have identified numerous ways to educate children about digital privacy and security. Quayyum et al. ~\cite{quayyum2021cybersecurity} summarized seven of these methods, spanning from training and warning to gamification. Prior work found digital privacy and security literacy training could improve children's awareness of online safety~\cite{desimpelaere2020knowledge}; however, such benefits are likely limited. Lastdrager et al.~\cite{lastdrager2017effective} found that traditional training approaches like anti-phishing training fall short in persisting children's privacy and security literacy. 
Given these insights, researchers have proposed novel educational interventions for digital privacy and security literacy specialized for children. 

Digital games (e.g., \cite{raynes2014gaming,maqsood2021design,kumar2018co}) and interactive applications (e.g., mobile apps~\cite{zhao2022koala, ge2024koala}, e-books~\cite{zhang2017cyberheroes,yap2020phone}) are among the most common approaches when developing privacy and security materials for children. These games and apps can target a broad range of digital privacy and security topics (e.g., \cite{kumar2018co, maqsood2021design}) or focus on just one topic (e.g., cybersecurity~\cite{quayyum2020cyber}). Through immersive and playful experiences, researchers have found these child-oriented approaches usually succeed in enhancing children's understanding of privacy and security concepts and awareness of risks in digital worlds \cite{maqsood2021design, zhang2017cyberheroes, yap2020phone}. Researchers have also explored how scenario-based gamification activities~\cite{blinder2024evaluating} can help children navigate privacy and security in different contexts and recognize that many situations do not have clear `right' and `wrong' answers.

Researchers have also criticized the shortcomings of existing digital privacy and security educational tools. In their systematic review of literature on cybersecurity education, Sağlam et al.~\cite{sauglam2023systematic} argued that most studies lack details on the timing of when content should be taught; this is especially important because children's experiences and abilities will vary significantly based on their age.
Other researchers have noted that current educational materials fall short in grounding privacy and security theories, making it hard to quantify what specific knowledge of privacy and security children take away from learning~\cite{kumar20225ds, kumar2023understanding}. 

The potential of in-school learning experience to promote children's privacy and security awareness has been highlighted in prior works~\cite{hartikainen2019children,lastdrager2017effective,livingstone2020data,rahman2020importance}. However, the vast majority of educational materials developed to help children learn about digital privacy and security are created for informal contexts like family-based learning~\cite{liu2024integrating, alghythee2024towards, ge2024koala}. In contrast, few approaches focused on digital privacy and security education within school settings.
Moreover, most approaches only support one-off learning in a transient period, while there are scarce tools that provide instruction and knowledge review. The present work builds a set of micro-lessons that covers four topics over four weeks and is scaffolded based on a child's age and grade. 

\subsection{Challenges to Teaching Digital Privacy and Security in Grades K--8} 
Many primary-level schools have employed technology in the classroom to facilitate teaching, learning, and social activities. With the rise of technology-integrated classrooms, CSCW researchers have identified digital privacy and security as a major concern in such sociotechnical settings \cite{wagman2023we} and have investigated teachers' perspectives of technology use at school including privacy and security concerns when using that technology~\cite{kumar2019privacy, lu2021coding, martin2023teacher, bacak2022elementary, chanenson2023uncovering}. Researchers have also explored social dynamics through digital privacy and security in the classroom, such as how teachers help children avoid online risks they encounter at or off school~\cite{maqsood2021they, martin2023teacher}, and privacy tensions between school/teachers and parents~\cite{kumar2018co, wagman2023we}.


Privacy and security are generally regarded as important components of technology education~\cite{livingstone2020data, rahman2020importance}. Researchers have evaluated in-classroom digital privacy and security education policy and materials, finding a lack of ready-to-use lessons provided by regional administrators or official departments~\cite{lamond2022sok, sauglam2023systematic, kumar2023exploring}. Researchers have also identified major gaps in digital privacy and security education in elementary schools~\cite{kumar2019privacy} and tensions around having to find educational resources without administrative support~\cite{smith2023educators}. Beyond that, numerous studies have identified a lack of professional development for teachers on how and what to teach children regarding privacy and security ~\cite{maqsood2021they, drader2022digital,kumar2019privacy}. 

When looking at the limited research on in-classroom digital privacy and security interventions, developed lessons tend to target older students (i.e., high schoolers~\cite{nicholson2021understanding} and college students~\cite{culnan2009online, vsvabensky2020cybersecurity}). 
Recent efforts to develop lessons for younger children have focused on a single, narrow topic, such as AI privacy~\cite{10.1145/3613904.3642460}, digital citizenship~\cite{li2023integrating}, and critical data literacy~\cite{bilstrup2022supporting}.
Moreover, while prior studies have evaluated the need for in-classroom digital privacy and security educational approaches by asking children's opinions~\cite{nicholson2021understanding} or reviewing existing resources and educational policies~\cite{kumar2023exploring, maqsood2021they}, there is a deficiency of teacher's perspectives on the demand for teaching children privacy and security at school. In our present work, we address this gap by using co-design with K--8 teachers to develop a series of micro-lessons that span digital citizenship, digital security, digital privacy, and critical data literacy.

\section{Methods Overview}

To develop digital privacy and security lessons for Grades K--8, we performed the study in two parts: a formative study (Study 1; Sections \ref{sec:formative} and \ref{sec:formativefindings}) and a co-design and evaluation study (Study 2; Sections \ref{sec:designeval} and \ref{sec:designevalfindings}). 
To ensure our lessons were appropriate for a diverse set of learners, we partnered with two public elementary schools in two US metropolitan areas, both of which are majority-minority schools. 
\begin{itemize}
    \item \textbf{PARTNER\_SCHOOL\_S1} (S1) is a PreK--5 public school in the Northeast US with 550-650 students. It has 64\% African American students and 28\% Latino students; 78\% of students qualify for free or reduced lunch \cite{Montpelier}.
    \item \textbf{PARTNER\_SCHOOL\_S2} (S2) is a K--8 public school in the Midwest US with 400-500 students. It has 88\% Black students, 64\% of students qualify as low income, and 15\% quality as diverse learners \cite{Montpelier}. 
\end{itemize}


For the formative study (Study 1), we completed interviews with 14 teachers at our partner schools, discussing children's digital privacy and security educational needs, as well as teachers' needs on digital privacy and security instruction and professional development. Findings from the formative study address RQ1, which also helped us identify the main criteria for lesson development. Referring to those findings, we then conducted a co-design and evaluation process (Study 2) to build and refine the micro-lessons shown in Fig. \ref{fig:studyprocedure}. We address RQ2 and RQ3 through two additional rounds of interviews with 13 teachers. All components of our study were approved by our Institutional Review Board (IRB). In the following sections, we describe both studies.  



\section{Study 1: Formative Study on Design Needs For Digital Privacy and Security Lessons}
\label{sec:formative}


\subsection{Study Procedure and Participants}
We conducted seven individual and small-group interviews with 14 PreK--8 teachers between February and April 2021. All sessions were conducted over Zoom and lasted 40-60 minutes. 

During each session, we guided a discussion with teachers on topics spanning: 1) an overview of what constitutes children's digital privacy and security literacy; 2) privacy, security, and safety concerns about children's technology use during and after school; 3) instructional approaches and curricular needs in digital privacy and security education; and 4) professional development experiences and needs in digital privacy and security education. Since this study was conducted while most schools were still engaged in remote learning, we also discussed privacy, security, and safety concerns and challenges associated with online instruction. See the full protocol in Section \ref{sec:formativeprotocol} of the Appendix.

Among the 14 participants in the formative study (\formative{T1--14}), nine worked at S1, while five worked at S2. There were two male and 12 female participants, which aligns with the gender skew of elementary school educators~\cite{nces2023publicteachers}. Our participants spanned a large range of grades and subjects, and we purposefully did not limit recruitment to certain subjects because technology is used frequently in schools across all subjects, and we felt that all teachers might have useful insights into how to structure content to best suit children's learning needs. See Table \ref{tab:demofocus} for demographic details.



\begin{table}[]
\caption{Participant Demographics of Formative Study}
\label{tab:demofocus}
\renewcommand\arraystretch{1.2}
\resizebox{\linewidth}{!}{
\begin{tabular}{lllllll}
\hline
\textbf{School ID}  & \textbf{Sessions}             & \textbf{PID} & \textbf{Gender} & \textbf{\begin{tabular}[c]{@{}l@{}}Grade(s)\\ Taught\end{tabular}} & \textbf{Subjects Taught}                    & \textbf{Role Note} \\ \hline
\multirow{9}{*}{PARTNER\_SCHOOL\_S1 (S1)} & \multirow{2}{*}{FSession \#1} & \formative{T1}           &     Female            & 4                                                                            & Math, Science                                        &                        \\
                    &                               & \formative{T2}           &      Female           & 4                                                                            & Math, Science                                        &                        \\ \cline{2-7} 
                    & \multirow{3}{*}{FSession \#2} & \formative{T3}           &       Female          & K                                                                            & Academic subjects (Not disclosed specifically)       &                        \\
                    &                               & \formative{T4}           &       Female          & K--3                                                                       & Math                                                 &                        \\
                    &                               & \formative{T5}           &        Female         & 1                                                                            & Academic subjects (Not disclosed specifically)       &                        \\ \cline{2-7} 
                    & \multirow{2}{*}{FSession \#3} & \formative{T6}           &         Female        & 3                                                                            & Math, Science                                        &                        \\
                    &                               & \formative{T7}           &        Female         & PreK--5                                                                    & Music                                                &                        \\ \cline{2-7} 
                    & \multirow{2}{*}{FSession \#4} & \formative{T8}           &      Female           & PreK--4                                                                    & All academic subjects                                &                        \\
                    &                               & \formative{T9}           &       Female          & K--5                                                                       & Information Literacy & Media Specialist       \\ \hline
\multirow{5}{*}{PARTNER\_SCHOOL\_S2 (S2)} & \multirow{2}{*}{FSession \#5} & \formative{T10}          &       Female          & K--8                                                                       & Information Literacy           & Librarian              \\
                    &                               & \formative{T11}          &      Female           & K                                                                            & Academic subjects (Not disclosed specifically)       &                        \\ \cline{2-7} 
                    & \multirow{2}{*}{FSession \#6} & \formative{T12}          & Male            & 2                                                                            & Academic subjects (Not disclosed specifically)       &                        \\
                    &                               & \formative{T13}          &      Female           & K--8                                                                       & Spanish                                              &                        \\ \cline{2-7} 
                    & FSession \#7                  & \formative{T14}          &     Male            & 4                                                                            & Math, Science                                              &                        \\ \hline
\end{tabular}}
\end{table}

\subsection{Data Analysis}
\label{sec:formativeda}

Interview transcripts were imported into the qualitative software analysis tool MaxQDA and analyzed through iterative, deductive open coding and thematic analysis, following the processes outlined by Salda\~na~\cite{saldana2021coding} and Braun \& Clarke \cite{braun2006using, braun2019reflecting}. First, one research team member constructed the initial codebook by extracting emerging structural codes and subcodes from all transcripts. Two other team members were then trained with the initial codebook, after which they finalized the codebook and performed another round of open coding. Every transcript was coded at least twice. The research team held regular meetings to discuss the emerging codes, update the codebook, and compare the analysis divergences to reach a consensus. Our final codebook included four structural codes related to designing privacy and security lessons: 1) digital privacy and security concepts for children; 2) privacy and security curriculum; 3) professional development; and 4) designing digital privacy and security micro-lessons. We present the final codebook in Table \ref{tab:formativecode} in the Appendix (Section \ref{sec:appcode}).

We then exported all coded excerpts to facilitate thematic analysis \cite{braun2006using}. Two team members read through all excerpts associated with a given subcode, identified patterns within the data, and wrote analytic memos for each subcode. During this process, the research team met regularly and reviewed, discussed, and revised these memos, eventually grouping overarching themes. In the following section, we present key themes regarding the current state of digital privacy and security in classroom education, as well as the design needs for developing digital privacy and security curriculum.
\section{Study 1: Formative Study Findings}
\label{sec:formativefindings}

Echoing prior works \cite{kumar2019privacy, maqsood2021they}, none of the teachers we spoke with received digital privacy and security curricula from their district, regardless of some programs teaching kids to keep safe in general. Some teachers (5/14) independently sought out or developed their own resources to teach the class. These teachers commonly mentioned using online teaching resources on digital privacy and security, such as Common Sense Media~\cite{commonsense_digital_citizenship} (\formative{T9}, \formative{T10}, \formative{T12}) and Nearpod~\cite{nearpod2023digitalcitizenship} (\formative{T2}), to cover basic concepts of digital citizenship, developing strong passwords, and exercising caution when following hyperlinks. One teacher (\formative{T14}) adopted resources on social and emotional learning, such as teaching tolerance online resources, to supplement digital privacy and security teaching. However, teachers raised concerns about the insufficient depth of these online resources. As \formative{T9} mentioned: \textit{``I think they don't get into the real meat of privacy and security beyond the surface.''}

Most teachers we spoke to received at least some professional development about digital safety. However, they received minimal to no professional development related to best practices for helping their students learn about digital security and privacy. As \formative{T12} summarized: \textit{``There's one on developing good passwords and not falling for phishing scams and things like that, but that's all for the teachers. It's not really how to teach the students.''} Reflecting on the relative lack of current privacy and security curriculum and professional development, teachers expressed a desire to enhance privacy and security learning in their schools. 

In the following section, we extend prior work \cite{kumar2019privacy} by presenting teachers' suggestions on the topics they considered important to be taught, as well as recommendations for designing in-classroom digital privacy and security lessons based on their experiences.

\subsection{Suggested Teaching Topics Around Digital Privacy and Security Literacy}
\label{subsec:need1}
When asked to define digital privacy and security literacy in their own words, teachers emphasized aspects of information sharing, digital literacy, and digital citizenship. Indicating online risks arising from specific incidents in the classroom, many teachers suggested topics that should be taught in digital privacy and security education. We expand teachers' statements regarding digital privacy and security literacy below.


\subsubsection{Appropriate Online Information Sharing And Privacy}
Many teachers (8/14) regarded digital privacy and security literacy as recognizing the difference between public and private information. Teachers expressed concerns about children oversharing private information online, which has been widely identified in prior work \cite{martin2023teacher, maqsood2021they, pir2023applying}. For example, \formative{T11} was worried about how her students post on social media: \textit{``Something was shared, a picture was taken and it was spread across (social media). And I don't think they understand once you post something it's pretty much there forever, and anybody has access to it.''} Teachers described children paying little attention to personal information privacy when logging in to school accounts on shared devices and forgetting to log out (\formative{T3}, \formative{T9}). Some teachers (4/14) spoke about promoting children's privacy and security awareness, with \formative{T11} emphasizing: \textit{``Keeping their digital footprint safe and making sure that they know how to keep their passwords and their sites safe. And then also knowing what they post online matters, it just doesn't disappear.''} Given that \textit{``technology is being used at younger and younger ages,''} \formative{T14} regarded keeping PII confidential as an increasingly critical aspect of privacy and security literacy, suggesting that children should know \textit{''how you're curating an identity on the Internet, like on social media.''} 

\subsubsection{Evaluating The Credibility Of Online Information for Safety}
Some teachers (7/14) defined privacy and security awareness as the ability to judge whether digital content is trustworthy---specifically, children's capacity to determine when it is safe to click on a hyperlink and evaluate the veracity of content found on websites, including embedded ads. As toxic speech and misinformation have become more prevalent on social media platforms, content-related risks, such as getting exposed to inappropriate content, have been identified as one of the most severe online risks for children~\cite{martin2023teacher, hasebrink2009comparing, maqsood2021they, pir2023applying}. \formative{T5} worried about children using Youtube: \textit{``YouTube is a helpful resource, but it is also a beast in itself. I wish that there was a way to [...] know exactly what these kids are on.''} 

Aligning with prior research~\cite{zhao2019make}, teachers also described younger children having a harder time verifying the veracity of online information in specific situations, such as when a site is \textit{``offering them a free game (that is possibly a scam), they might click on it''} (\formative{T12}). Some teachers acknowledged that this aspect of digital security and privacy is challenging even for adults, with \formative{T10} noting, ``\textit{The sophistication of scams and phishing, the landscape has become more difficult. How do you help children recognize potential harm, when it's hard for you as an adult to see it as well?''} Therefore, several teachers suggested that an important aspect of privacy and security literacy for children is helping them learn when to seek help and what to do when they make a mistake. For example, \formative{T13} suggested that children should be equipped to both distinguish unsafe information and respond to potential threats, seeking help if necessary: \textit{``Always checking with an adult. We’re there to support them if they don't feel safe. Or if they’re on their own and they see something that's strange or they don't feel comfortable, they should just log out. ''}  

\subsubsection{Being A Good Digital Citizen}
Some teachers (6/14) spoke of digital privacy and security literacy in terms of respectful and responsible participation in online communities and activities. Prior studies have identified online risks for children, most often cyberbullying~\cite{martin2023teacher, hasebrink2009comparing, maqsood2021they, bacak2022elementary, pir2023applying}. Similarly, some teachers we spoke to elaborated on how toxic online activities could hurt children. \formative{T4} emphasized how anonymity could taint interactions and lead to various forms of cyberbullying. She also treated it as an important concept to be taught: \textit{``Another thing I feel is really important to teach children... is what the effects of anonymity can do to people, and how people will act very differently when they know that they're posting anonymously rather than posting with their name attached to something.''} Teachers wanted their students to understand how their interactions online could have a broader impact on their own and others’ emotional safety and well-being, with \formative{T13} framing it as, \textit{``we're going to be respectful, we're going to be gentle and kind to each other and to ourselves.''} Teachers also stressed the importance of learning how to effectively regulate behavior when confronted with deceptive and manipulative design features, such as continuous scrolling and game incentive systems. \formative{T14} described wanting to teach children to \textit{``understand how design can influence you and how you can push back against those things if you're not liking the way you're being influenced.''}

\subsection{Recommendations and Needs for Future In-Classroom Digital Privacy and Security Lessons}
Teachers highlighted the importance of building students’ conceptual understanding and applicable skills related to different aspects of digital privacy and security. They also spoke about other desires for lessons and professional development for teaching these topics to children. In this section, we describe teachers' recommendations and needs for digital privacy and security education at school.

\subsubsection{Differentiate Between Educational Goals and Needs Across Different Grade Bands}
\label{subsec:need2}
Upper-grade teachers emphasized challenges with students effectively and ethically conducting online activities. They felt a digital security and privacy curriculum should have practical activities to help students learn appropriate behaviors in real implementations, such as \textit{``experiment[ing] with social media platforms so that they know how to use them appropriately''} (\formative{T14}). 

While teachers in upper-grade bands focused more on issues related to actual uses of digital devices and advanced concepts, teachers working with younger children (grades PreK--2) expressed a desire to support children's learning in a way that was appropriate to their developmental stage cognitively and physically, since for younger children, \textit{``this is their first time really having the Internet and a computer for a long period.''} (\formative{T12}). 
Teachers perceived young children's developing abilities in reading, writing, and cognition as critical challenges in privacy and security learning. \formative{T8} shared: \textit{``At this age since the children aren't even reading yet... they end up in the wrong place because they don't know what they're doing.''} They thus requested materials that would be suitable for pre- and emergent readers struggling to make sense of text-centered navigation in digital environments.

\subsubsection{Connect Privacy and Security to Everyday, Relatable Contexts for Children}
\label{subsec:need3}
Building on our last example of considering how to help young children understand complex concepts, some teachers (5/14) also suggested ways to connect lessons to children's daily lives. They envisioned the examples could be either metaphors for children's experiences or analogies from others' experiences. \formative{T14} described an example from her teaching: \textit{``I told the students from my own life, `I have my work email and my personal email, just like you have your school email and then many of you have Roblox accounts, which are your personal accounts...' And I told them, `I would never use my work email to get in contact with my friends about something.'''} Moreover, the concepts of privacy and security are ambiguous and difficult for young children to understand. Therefore, teachers spoke of a need for clear examples that would be comprehensible to students less adept in abstract reasoning, \textit{``a concrete scenario where if this happens, this could be a concern, and here's why''} (\formative{T7}). 

\subsubsection{Create Robust and Easy-to-Integrate 'Mini-Lessons'}
\label{subsec:need4}
The format is as important as content in determining the types of curricular resources that would be most helpful in addressing teachers’ digital security and privacy needs. Some teachers expressed a need for content that could be easily shared with families, as \textit{``they may not know what [privacy] is and how to help their child''} (\formative{T2}). In addition, teachers wanted resources to flexibly integrate privacy and security content into their regular instruction plans. In terms of curriculum, teachers emphasized the importance of having some autonomy in selecting the activities, instructional sequence, and format that would best support their students' needs. Some teachers (6/14) advocated developing `mini-' or `micro-' lessons that were both flexible to implement and easy to distribute. For example, \formative{T11} perceived mini-lessons as a reference to guide and tailor privacy and security teaching in the classroom: \textit{``Maybe some very short mini-lessons. Not even like, here are all the things that you have to do, but here are some suggestions if you run into these kinds of issues. Here's a quick little mini-lesson that you can do with your class to help mitigate that issue.''} Many teachers also felt a digital format --- \textit{``having some kind of a resource that's housed in a findable place, and that is hyperlinked to those mini-lessons''} (\formative{T10}) --- would facilitate distribution and engagement throughout their broader classroom and school communities.

\subsubsection{Implement Ongoing Privacy- and Security-Focused Professional Development} 
\label{subsec:need5}
Some teachers (6/14) expressed a desire for ongoing professional development and support with built-in opportunities to apply and practice what they’d learned. Many teachers felt that a single workshop would prove inadequate in preparing them to effectively teach their students about digital security and privacy. As \formative{T8} explained: \textit{``We've had lots of digital training, but it's an hour here, this is what you got, go. The actual time available to be able to work with it and play with it and try to develop it, isn't there [...] You can't just hand it to me and say, `Here, you have to let me work with it and involve me in it.'''} Therefore, teachers widely acknowledged their need for more training. Moreover, teachers suggested a more interactive and practical approach toward professional development. For instance, \formative{T9} suggested having a platform where teachers could develop and refine materials and receive feedback on their implementation: \textit{``Having something where we're able to check in or where someone can follow up and say, `I observed this. Instead of doing that, this is how you could actually implement this,' or `this is how you could actually integrate technology at this point.'''}

\section{Study 2: Micro-Lessons Co-Design and Evaluation}
\label{sec:designeval}


Overall, the formative study (Section \ref{sec:formativefindings}) clarifies a set of factors teachers felt were important for developing learning programs for K--8 children that cover digital privacy and security. Based on these findings, we identified five primary needs for designing lessons on digital privacy and security for K-8 classrooms:

\begin{itemize}
    \item Need \#1: Children need to learn various skills regarding digital privacy and security, including but not limited to appropriate information sharing, evaluating the credibility of online information, and being good digital citizens (Section \ref{subsec:need1}).
    \item Need \#2: Curricula for different grade bands should have different goals and lesson formats when learning the same topic, and be considerate of children's current knowledge, skills, and capabilities (Section \ref{subsec:need2}).
    \item Need \#3: Using everyday, relatable contexts and examples could help concretize abstract concepts of digital privacy and security (Section \ref{subsec:need3}).
    \item Need \#4: Digital privacy and security curricula should be short and flexible to fit into daily teaching schedules, yet robust and easily shareable with families (Section \ref{subsec:need4}).
    \item Need \#5: Digital privacy and security curricula should include simple and flexible resources to equip teachers with privacy and security background knowledge before teaching (Section \ref{subsec:need5}).
\end{itemize}

\begin{figure}
    \centering
    \includegraphics[width= \textwidth]{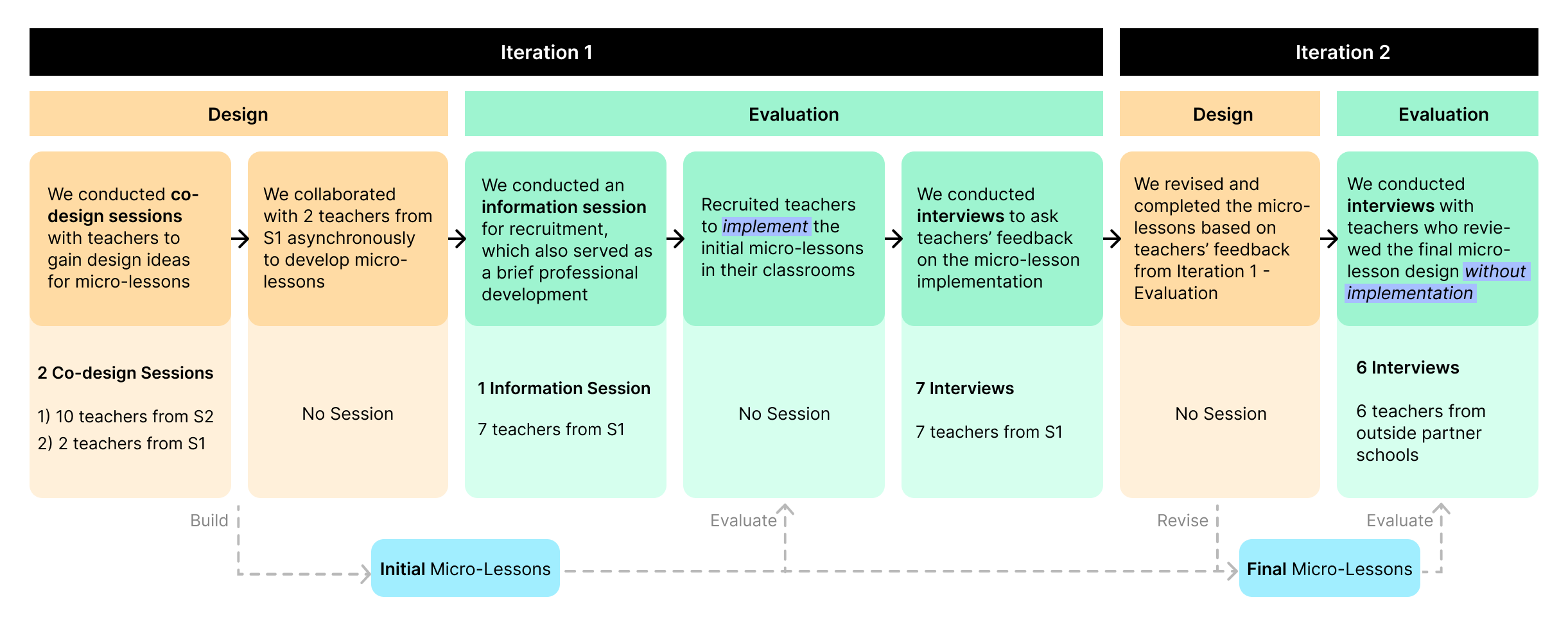}
    \caption{Study procedure of micro-lessons design and evaluation}
    \label{fig:studyprocedure}
\end{figure}

Using these findings as a guide, we developed digital privacy and security micro-lessons through a co-design and evaluation process, as shown in Fig. \ref{fig:studyprocedure}. In the first iteration, the research team conducted co-design sessions with 12 teachers from our two partner schools to create content for the initial lesson framework. Using ideas generated in these sessions, we developed a micro-lesson outline and a sequence of micro-lesson activities in conjunction with two lead teachers at S1. Once the initial micro-lesson framework was finalized, we conducted a professional development session with teachers at S1 to introduce and describe how to use the micro-lessons. Seven teachers at S1 then implemented the micro-lessons in their own classrooms, then participated in one-on-one interviews with the research team to share feedback and suggestions on further improving the micro-lessons. After revising the micro-lessons based on this feedback, we then completed an additional six interviews with outside teachers to evaluate the finalized micro-lesson content.


\subsection{Design and Evaluation Iterations}
\subsubsection{Iteration 1: Initial Micro-Lessons Design}
\

\begin{figure}
\centering
\begin{subfigure}{0.39\textwidth}
  \centering
  \includegraphics[width=1\linewidth]{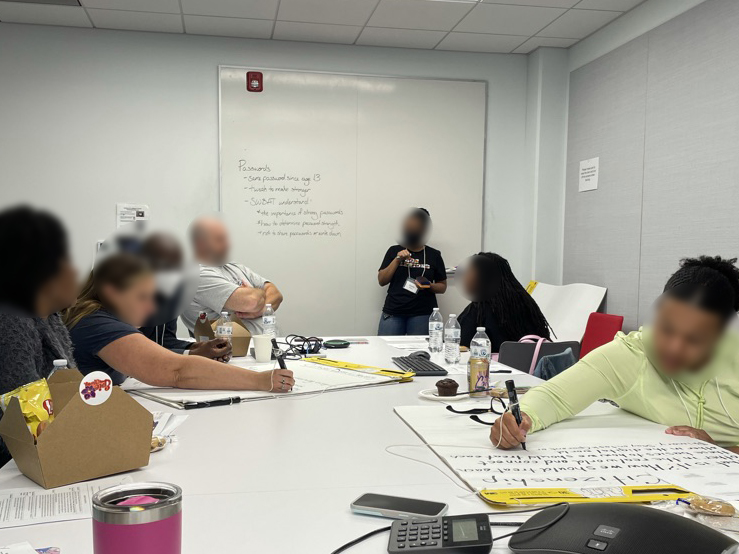}
  \caption{Teachers brainstorm content ideas during the co-design session.}
\end{subfigure}
\hfill
\begin{subfigure}{0.2925\textwidth}
  \centering
  \includegraphics[width=1\linewidth]{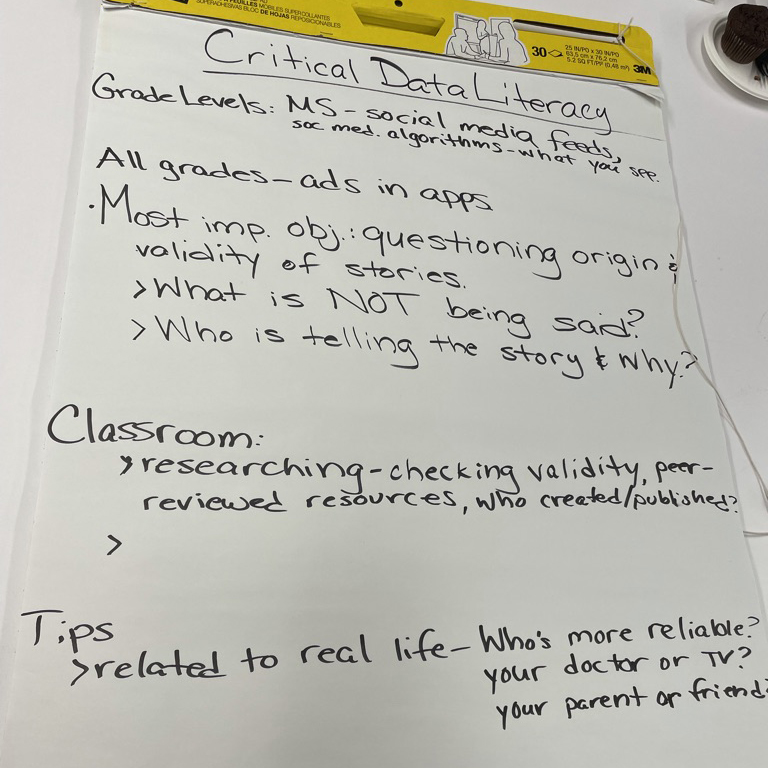}
  \caption{Notes on teaching critical data literacy made by teachers.}
\end{subfigure}
\hfill
\begin{subfigure}{0.2925\textwidth}
  \centering
  \includegraphics[width=1\linewidth]{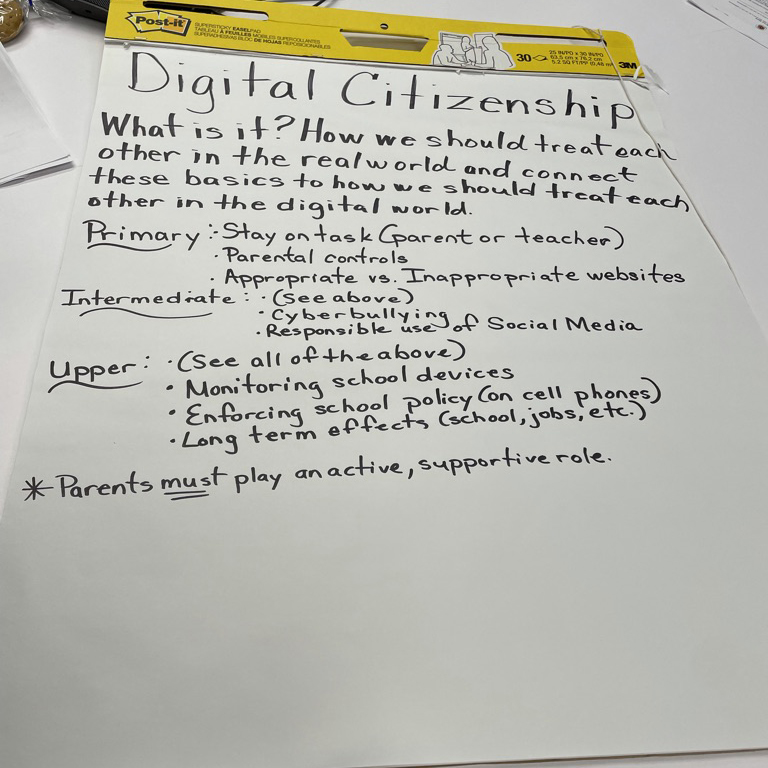}
  \caption{Notes on teaching digital citizenship made by teachers.}
\end{subfigure}
\caption{Photos of the co-design session held in S2.}
  \label{fig:codesign}
\end{figure}

\textbf{Co-design sessions.} 
We conducted two in-person co-design sessions with teachers to create the initial micro-lessons. First, we held a half-day co-design session with 10 teachers at S2 in September 2022 (see Fig. \ref{fig:codesign}). Participating teachers taught a range of subjects, from mathematics to physical education and counseling. None of the teachers who joined this co-design session participated in our formative study (Study 1). We provided teachers with a basic overview of three starter topics drawn from privacy and security concepts and practices that arose in the formative study: how to set and manage strong passwords; being a good digital citizen; and critical data literacy. We asked teachers to reflect in small groups based on grade bands and subject areas on these topics and what concepts within these topics would be appropriate for students to learn. Next, we shared examples of privacy and security resources for teaching these concepts (e.g., Common Sense Media, Google's `Be Internet Awesome' program, and academic research).
We asked teachers to discuss how these resources could be curated for teaching their students about privacy and security in the context of their subject areas. The session was audio-recorded and we took notes on poster boards to summarize the group discussions and reflective aspects of the workshop. We then extracted and organized design ideas from audio recordings, teacher breakouts, and notes.

In January 2023, we conducted a similar, but shortened, co-design session with T6 and T9 from S1. During this session, we shared the ideas for micro-lessons that emerged from the first session with S2, then created a plan to finalize and implement the micro-lessons with teachers at S1. We worked asynchronously with T6 and T9 over the next several weeks in a shared Google Document to form the structure and activities for the initial micro-lesson framework.

\begin{figure}
    \centering
    \includegraphics[width= \textwidth]{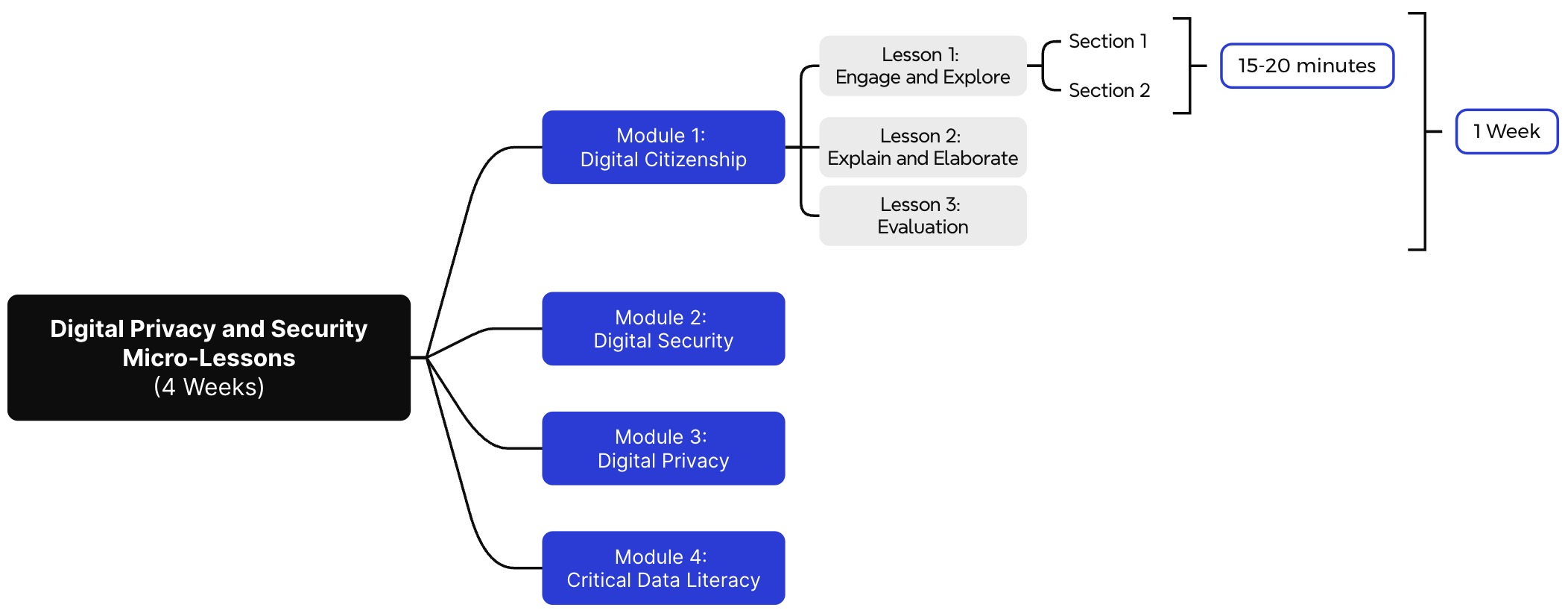}
    \caption{Structural components of our micro-lessons.}
    \label{fig:lessonstructure}
\end{figure}

\textbf{Design decisions and initial design overview.} The micro-lessons structure is displayed in Fig. \ref{fig:lessonstructure}. The initial micro-lessons consist of four modules to be implemented over four weeks, each covering one topic related to digital privacy and security. We created this lesson structure to address Need \#1 regarding the types of skills teachers wanted children to develop. The first module (Digital Citizenship) introduces digital literacy and cyberbullying. The next two modules (Digital Security; Digital Privacy) cover the basics of appropriate information sharing and evaluating the credibility of online information, as well as other concepts like password security. The final module (Critical Data Literacy) explores topics like how the information pipeline works online. 


To address Need \#4 (keeping micro-lessons compact), each module has three 15-20-minute lessons that can be taught over three days in a given week. We utilized the 5E instructional model~\cite{bybee2006bscs} suggested by the lead teachers at S1 and consisting of five stages for learners to become familiar with a topic: engage, explore, explain, elaborate, and evaluate. Our initial micro-lessons followed this format to help children engage in and explore the topic in the first micro-lesson, explaining and elaborating the concept in the second micro-lesson, and evaluating children's learning outcomes in the third micro-lesson. Each micro-lesson contains two sections of activity and may include watching topic-themed videos, open-ended discussions, digital educational games, and/or brief assessments. 

\begin{figure}
    \centering
    \includegraphics[width= \textwidth]{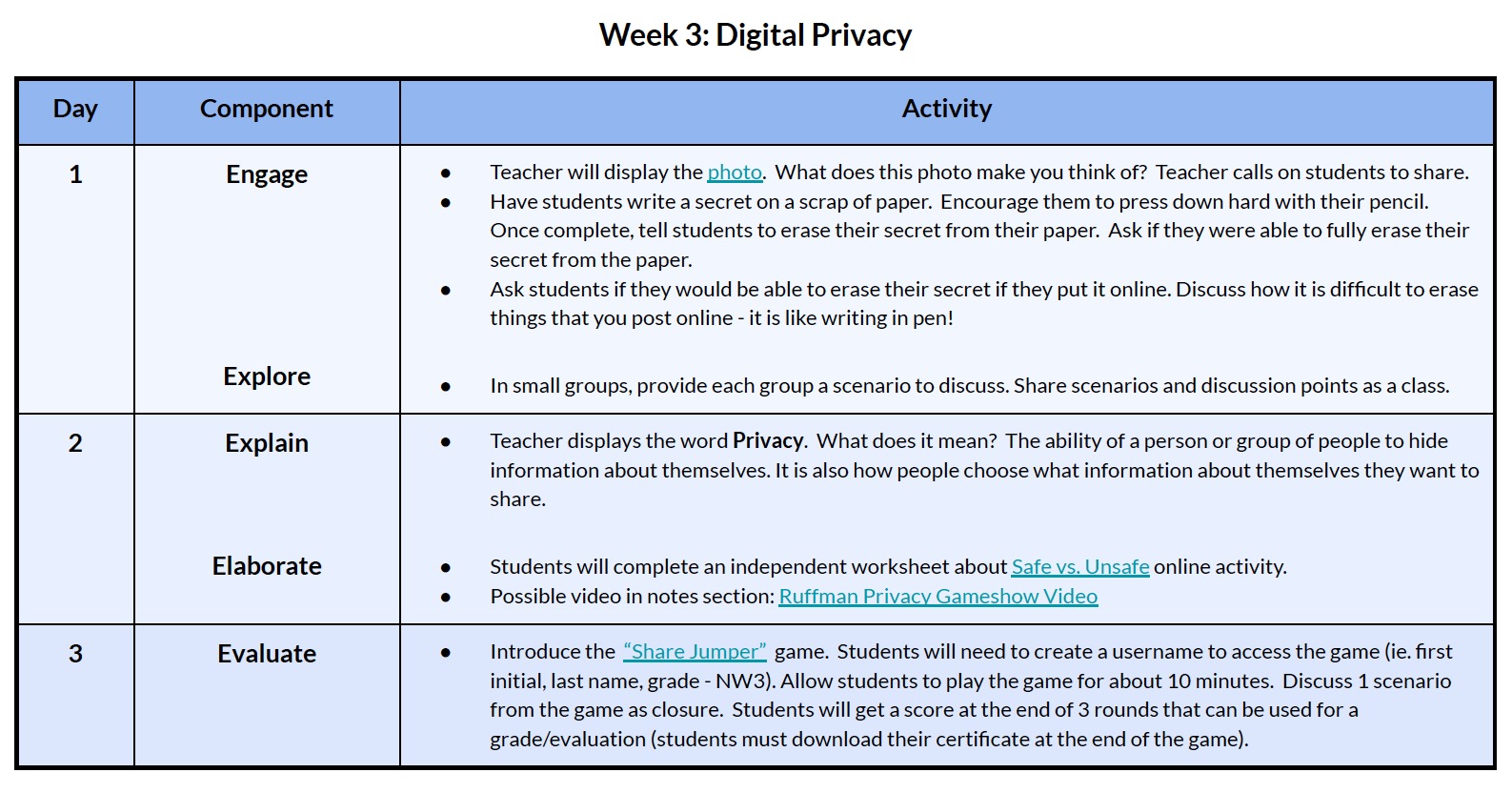}
    \caption{A work-in-progress design of study module 3 (Digital Privacy) when developing initial micro-lessons.}
    \label{fig:initiallesson}
\end{figure}

To address Need \#3, all activities were selected, designed, and curated to connect to children's real-life experiences of privacy, security, and safety online and offline. In response to Need \#5, a brief instructional guide for each lesson was provided in the initial lesson framework to help teachers understand the purpose of each lesson and to introduce privacy and security concepts to teachers. We did not address Need \#2 in this stage of the iterative design process, since we intended to better capture the learning needs across different grade bands through implementation and evaluation. In Fig. \ref{fig:initiallesson}, we present a work-in-progress design for \textit{Module 3 (Digital Privacy)} when developing the initial micro-lessons, where the research team and collaborated teachers brainstormed lesson contents, activities, and instructional guides fitting in the lesson structure.

\subsubsection{Iteration 1: Initial Evaluation Via Classroom Implementation}
After developing the initial micro-lessons with T6 and T9, we held a professional development session for seven teachers from S1 in March 2023 to go over the micro-lesson content and teacher implementation plans. 
When the teachers finished their micro-lesson implementation, which took between 1-4 weeks depending on how many modules they implemented, we scheduled a 60-minute semi-structured interview with each of them to discuss their experience with and feedback on the micro-lessons. All interviews, led by the second author, were conducted remotely via Zoom between April and June 2023. During the interview, we asked participants' opinions on the micro-lesson design, implementation experiences and challenges, and additions or adaptations to the materials. See Section \ref{sec:evaluationprotocol} in the Appendix for the full protocol.


\subsubsection{Iteration 2: Iterative Revision of Micro-Lessons}
Following the completion of Iteration 1, we revised the micro-lessons based on explicit feedback provided by teachers who implemented the micro-lessons in their classes, as well as design needs we did not fully address in the initial design. 

\textbf{Teacher feedback from initial evaluation.} Below, we summarize teacher feedback from the initial evaluation on how the micro-lesson design could be further improved.

\begin{itemize}
    \item Feedback \#1: Teachers from the different grade bands shared their experiences with the curriculum, highlighting the varied needs, abilities, and challenges of children from these grade bands for privacy and security learning. They suggested more differentiation according to grade bands in the lesson plans. 
    \item Feedback \#2: Teachers found it challenging to manage the lesson progress during class and lesson duration, for which they suggested a consistent structure for every lesson to make the learning process more manageable and adjustable.
    \item Feedback \#3: Teachers struggled with limited knowledge and teaching experience in privacy and security when preparing and teaching the lesson, for which they called for more detailed, concrete instructional guidance.
\end{itemize}

\textbf{Summary of design revisions.} 
First, to incorporate Feedback \#1 (which also reflected Need \#2 from the formative study), we broke down each lesson into content tailored based on grade bands (e.g., K--2, 3--5, 6--8). The lessons also scaffold, with each activity building on knowledge learned from prior micro-lessons on that topic, which allows the materials to scale as children advance in school. For each lesson, each grade band has unique lesson content that is tailored to the learning ability and needs of children at that age. We also adjusted and redesigned grade-appropriate activities for Grades K--8. 

Then, to address Feedback \#2, we standardized the format for each lesson and simplified the terms of the 5E framework that we used before. Each module has a unified agenda with a video and discussion section plus an activity section (the first and second lesson of each module) or a reflection section and an activity section (the third lesson of each module). This change creates a more consistent and manageable schedule for teachers. 
We added additional videos and activities where needed to make the final micro-lessons comprehensive and provide material for each grade band. Similar to the initial lesson framework design, when developing material to construct our micro-lessons, we either referred to resources from open-source media and literature or used resources that were created by the research team.

Building upon the initial instructional guides and introductory materials on privacy and security concepts, we added support for teachers in the final micro-lesson instructional document in response to Feedback \#3. These included step-by-step guides on each activity's purpose and goals, as well as any steps needed for preparation and a brief background on the privacy and security concepts being covered. We also added guiding questions, class reflection questions, and teacher resources for each lesson. 


\textbf{Final design overview.} The final micro-lessons cover the same core concepts as in the initial design (Fig. \ref{fig:lessonstructure}). 
Each of the four modules is crafted to be self-contained and independent, while lessons within modules sequentially build upon one another. Learning tasks are organized into three 15-20-minute lessons and allow teachers some flexibility to adjust timing based on their personal teaching schedule. 


Following the 5E framework, each module is structured as follows: The first micro-lesson provides an overview of the module topic. The second micro-lesson is an advanced lesson to explore complex concepts of the topic and connect the concepts with everyday online activities. 
The third micro-lesson provides a wrap-up and review, with the goal of having children apply their understanding to related online activities. The first two lessons of each module contain a video \& discussion section (5 minutes), where children watch a short video related to the covered topic and then reflect on questions provided in the micro-lesson instructional document led by the teachers; and an activity section (10-15 minutes), where the teachers conduct a digital or analog activity with children following the instructions in the micro-lesson instructional document. In the third lesson, the video \& discussion section is substituted with a reflection section for students to revisit what has been covered in the prior two lessons. 
The outline of the first lesson in \textit{Week 1 (Digital Citizenship)}, which represents the structure of most of the lessons, can be found in Fig. \ref{fig:finallesson1} and Fig. \ref{fig:finallesson2} in Appendix (Section \ref{sec:finaldesign}). The complete micro-lessons document can be downloaded at \url{https://spe4k.umd.edu/wp-content/uploads/2024/06/Connecting-Contexts-Lesson-Plans-May-2024.pdf} and an overview is shown in Table \ref{tab:final_micro-lessons}.

\begin{table}
    \centering
    \caption{Overview of finalized micro-lessons. }
    \label{tab:final_micro-lessons}
    \renewcommand\arraystretch{1.2}
    \resizebox{0.85\linewidth}{!}{
    \begin{tabular}{p{0.15\linewidth}p{0.3\linewidth}p{0.46\linewidth}}
    \hline
    \textbf{Module Topic} & \textbf{Main Goal} & \textbf{Micro-Lesson Details}  \\ 
    \hline
      Module 1: Digital Citizenship  &  Learn, interact, and share online in a way that matches personal values. & Learn the basics of what the Internet is, how it is used, and responsible use of technology. Explore similarities and differences between being a good citizen online and in person. Examples of harmful online behaviors are introduced. Reflect on accomplishments and goals as digital citizens. \\ \hline
       Module 2: Digital Security & Learn ways to protect themselves and others from risky online situations. &  Discuss safe online behavior, including online tracking and strong passwords. Learn to evaluate the risks of content viewed, shared, and clicked to deepen comprehension of responsible online conduct. Create goals for staying safe and secure online. 
       \\ \hline
       Module 3: Digital Privacy & Learn how to weigh trade-offs of sharing different types of information in online and offline contexts. &  Define digital privacy and apply understanding to own lives. Enhance understanding of online privacy. Learn how to maintain digital privacy and evaluate privacy values. \\ \hline
        Module 4: Critical Data Literacy & Learn how digital applications and games collect and use personal information. & Understand what data is and about different forms of data collection. Explore why companies collect data and the potential outcomes of data collection. Reflect on all four modules to develop sustainable practices and share lessons learned with others. \\ 
        \hline
    \end{tabular}}
\end{table}

\subsubsection{Iteration 2: Evaluation Interviews with Outside Teachers}
To evaluate the finalized micro-lessons, we extended our participant pool beyond our partner schools to six elementary school teachers recruited from our national networks. Due to time, geographical, and collaboration constraints, these teachers were not able to implement the micro-lessons in their classrooms. However, interviews with these teachers included an in-depth review of the updated micro-lesson plans and garnered their perspectives on the micro-lessons. While teachers envisioned the micro-lessons in their schools hypothetically, their insights enabled us to: (1) see if the findings/needs we observed with our participants from our two partner schools were confirmed beyond those two locations and (2) obtain teachers' perspectives on how the resources would work with their specific teaching contexts in mind (i.e., how they interpreted the modifications we made and whether any aspects of resources were confusing or not feasible). This approach of gathering feedback from teachers prior to implementing has been found to be important both for collecting teachers' unique perspectives before major implementations (e.g., \cite{norooz2015bodyvis}) and for promoting teachers' learning and professional development \cite{goldman2022collaborative, kyza2017co}, which we deem important for teachers' exposure to privacy and security in the classroom \cite{kumar2017no}.\footnote{Owing to an administrative change at S2 in 2023, we were unable to recruit teachers from that school to evaluate the finalized micro-lessons.} 
All interviews were conducted between November 2023 and January 2024 virtually via Zoom and lasted about 30--45 minutes. The first author led all interview sessions. At least three days before the scheduled interview, we sent the micro-lesson instructional document to participants and asked them to provide initial comments by filling out a pre-interview survey, which also asked for their demographics and consent for the study. Before the interview started, we double-checked if the teachers had reviewed the document, ensuring they were familiar with the content of the micro-lessons. The topics covered in the interviews were similar to the previous interviews for the initial evaluation, but asked teachers to imagine themselves implementing the micro-lessons with their students since they did not have time to do the implementation. 

\subsection{Evaluation Session Participants}
We recruited participants for each phase of the micro-lesson evaluation separately. For the initial evaluation, we recruited seven teachers from S1 (\implementer{T5}, \implementer{T6}, \implementer{T9}, \implementer{T15--18}). Notably, three participants (\implementer{T5}, \implementer{T6}, \implementer{T9}) also participated in the formative study (Study 1). For the final evaluation, we recruited six US-based teachers through the research team's professional networks and institutional mailing lists (\commenter{T19--24}), including three teachers from public elementary schools, two from private elementary schools, and one teaching principal at a public elementary school. Participants had diverse backgrounds in grades taught, subjects taught, and teaching experience (7--26 years teaching). All participants were working with K--5 students at the time of the study, although two teachers had previous experience teaching children in grades 6--8. 
See Table \ref{tab:demointerview} for demographic details.


\begin{table}[]
\caption{Participant Demographics of Evaluation Study}
\label{tab:demointerview}
\renewcommand\arraystretch{1.2}
\resizebox{\linewidth}{!}{
\begin{tabular}{llllll}
\hline
\multirow{8}{*}{\textbf{\begin{tabular}[c]{@{}l@{}l@{}}Initial \\Evaluation\\ (Implementing) \end{tabular}}} & \textbf{PID} & \textbf{Gender} & \textbf{\begin{tabular}[c]{@{}l@{}}Implementation Grade(s) \\ (Other Grades Taught)\end{tabular}} & \textbf{Subject Taught}                    & \textbf{Role Note}       \\ \cline{2-6} 
                                  & \implementer{T5}*          &       Female          & 5                                                                                                   & Science, Social Studies, Health                      &                              \\
                                  & \implementer{T6}*          &       Female          & 4 (1)                                                                                               & Science, Social Studies, Health                      &                              \\
                                  & \implementer{T9}*          &       Female          & K--5                                                                                              & Information Literacy  & Media Specialist \\
                                  & \implementer{T15}          &      Female           & 3                                                                                                   & Math, Science, Health                                &                              \\
                                  & \implementer{T16}          &       Female          & K                                                                                                   & All academic subjects                                &                              \\
                                  & \implementer{T17}          &       Female          & 4                                                                                                   & Science, Social Studies, Health                      &                              \\
                                  & \implementer{T18}          &       Female          & 2                                                                                                   & All academic subjects                                &                              \\ \hline
\multirow{7}{*}{\textbf{\begin{tabular}[c]{@{}l@{}l@{}}Final \\ Evaluation\\(Reviewing)\end{tabular}}} & \textbf{PID} & \textbf{Gender} & \textbf{Grade(s) Taught}                                                                  & \textbf{Subject Taught}                    & \textbf{Role Note}       \\ \cline{2-6} 
                                  & \commenter{T19}          & Female          & 3--5                                                                                              & Computer Science                                     &                              \\
                                  & \commenter{T20}          & Male            & K--2                                                                                              & All academic subjects                                & Principal                    \\
                                  & \commenter{T21}          & Female          & 1                                                                                                   & Language Arts, Math, Social Studies, Science        &                              \\
                                  & \commenter{T22}          & Female          & 3                                                                                                   & Math, Reading, Writing, Science, Social Studies      &                              \\
                                  & \commenter{T23}          & Female          & 1                                                                                                   & Math, Reading, Science, Language Arts, Health        &                              \\
                                  & \commenter{T24}          & Female          & K--2                                                                                              & Computer Science               &                              \\ \hline
\end{tabular}}
\begin{tablenotes}
  \small
  \item *T5, T6, and T9 also participated in Study 1. Note that T5 and T6 changed grades and subjects since the formative study.
\end{tablenotes}
\end{table}

\subsection{Data Analysis}
All co-design sessions and interviews were video and audio recorded. In the co-design process, owing to the need for rapid movement between co-design sessions, we analyzed the audio files and transcribed big poster notes and field notes to help design the initial micro-lesson ideas. Our main analysis reported in the paper is the qualitative analysis of interview data from the 13 teachers participating in the evaluation study. Similar to the formative study, we used iterative, deductive open coding and thematic analysis \cite{saldana2021coding,braun2006using,braun2019reflecting}. First, we conducted open coding for interview transcripts to create an initial codebook. Here, one team member individually coded two transcripts and extracted emerging structural codes and subcodes. Another team member was then trained with the initial codebook and helped finalize the codebook. Our final codebook includes four structural codes: 1) class implementation approaches; 2) strengths and challenges of teacher teaching; 3) opportunities and difficulties in student learning; and 4) potential lesson improvement. We present the final codebook in Table \ref{tab:evaluationcode} in the Appendix (Section \ref{sec:appcode}).

Every transcript was then coded twice independently by two researchers. The research team held regular meetings to discuss emerging codes, update the codebook, and compare analysis divergences to reach a consensus. We then exported excerpts for each code and subcode for second-round thematic analysis, where the same team members read and re-read excerpts to identify patterns in the data, and then wrote analytic memos for each subcode. Via full-team discussions, we coalesced on three themes: 1) incorporating micro-lessons in elementary school infrastructure, 2)  micro-lessons teaching and learning experiences, and 3) gains from micro-lessons.



\section{Study 2: Evaluation Interviews Findings}
\label{sec:designevalfindings}



\subsection{Facilitators, Barriers, and Adjustments for Privacy and Security Micro-Lessons}
\label{subsec:DFinfrastructure}

According to teachers who participated in the evaluation interviews, micro-lessons could easily integrate into elementary school infrastructure for in-classroom privacy and security education. However, they also identified barriers that constrained how micro-lessons can be taught and adjustments needed to overcome these issues. 

\subsubsection{The Micro-lesson Format Helps Elementary School Teachers Discuss Privacy and Security Concepts With Children} 
\label{subsubsec:DFinfrastructureFacilitator}
Overall, most of the teachers we spoke to (iteration 1: 6/7; iteration 2: 5/6) found the micro-lessons easy to implement because of their clear lesson structure and guidance, flexible lesson format, and the abundance of linked educational resources about privacy and security. 
Specifically, the integrated and accessible resources on privacy and security topics decreased the amount of time teachers needed for class preparation. As \implementer{T5} shared: \textit{``Having a lot of the links set up in this, I didn't have to try to get a link to work or get something else so it flowed. I could copy and drop it into my slides for the day and just run with it.''}

Most teachers felt the micro-lessons were clear and well-structured and appreciated that the instructional document included step-by-step instructions on how each class should be implemented. For example, (\implementer{T17}) said, \textit{``Even if I picked and chose and went online and found stuff, it was nice to have a guide and a trajectory.''} These instructions allowed teachers to \textit{``just flow naturally from one activity to the next''} (\implementer{T6}). Likewise, \commenter{T19} said, \textit{``A teacher could pick up and just use it immediately and not have to do much background work,''} which they perceived as particularly valuable since teachers had limited time for class preparation. 
Teachers also appreciated the lessons were structured similarly to other academic STEM lessons (\implementer{T15}). Owing to the clear instructions and structured approach to doing each micro-lesson, one teacher (\implementer{T5}) asked an intern to lead the lesson because she believed that even a teaching novice could implement the micro-lessons. 

Teachers appreciated the flexibility micro-lessons offered to incorporate digital privacy and security learning into existing curricula. Several teachers felt this was especially helpful given that privacy and security concepts come up in multiple subject areas, such as social and emotional learning (\implementer{T5}, \commenter{T20}). Some teachers who implemented the lessons tried to integrate micro-lesson content into special topic classes. For instance, \implementer{T5} reported that she \textit{``put (the content of micro-lessons) into a quick social-emotional learning lesson.''}

Teachers who provided feedback on the finalized micro-lessons appreciated their short length, stating that one advantage of the micro-lessons was that teachers could easily integrate lessons into a normal class schedule instead of requiring separate sessions for these topics. (\commenter{T19}) noted that \textit{``a teacher could do it easily within a morning meeting and spend some of these that you could really break it down and do it in five or 10 minutes,''} while \commenter{T20} felt that the shorter lessons were less burdensome for teachers: \textit{``That gives teachers some flexibility to integrate these into the (existing) curriculum so that it doesn't feel like I have to stop my math lesson, or stop my reading lesson, to do four more things.''} \commenter{T24} further noted that the short length could keep children on track: \textit{``Their attention span. They say, the number of minutes is equal to your age, so if you're seven, seven minutes attention. That's why the 15 to 20 minutes was appealing to them.''}


\subsubsection{Infrastructural Difficulties and Class Planning Solutions}
\label{subsubsec:DFinfrastructureDifficulty}

Many teachers (iteration 1: 4/7; iteration 2: 5/6) raised concerns about potential infrastructural constraints to implementing micro-lessons, such as school policies on class time and digital devices. In response to these constraints, teachers actively adjusted their classes to better fit the micro-lesson contents into current elementary school educational infrastructure. While teachers liked the short format of micro-lessons to fit within a 15-20 minute time frame, they still identified challenges and constraints with implementation.

Elementary schools commonly have class timing constraints that limit teachers' ability to modify lessons; \commenter{T21} encapsulated this by saying, \textit{``[With] the requirements from the county where I work, time is always a crunch.''} However, the timing of micro-lessons was often a concern during classroom implementation. Teachers who implemented the initial lessons 
said that in some cases, the micro-lessons went significantly over the 15-20 minutes specified in the instructional document. When teachers prepared for their teaching, the uncertainty of how much time children would take to fulfill the tasks and understand the concepts was a huge factor in planning. \implementer{T18} expressed her worries about preparing for the \textit{Digital Privacy} classes: \textit{``I don't know if I fully got to play the Ruffman privacy game show video. I previewed it, but I didn't have as much time and I just wanted to make sure they understood how to keep your privacy to yourself. ... most of the time I try to get it all done in one day just because of everything else going on.''} Teachers who provided feedback on the finalized micro-lessons also expressed concern about the micro-lesson length. \commenter{T24} suggested that the micro-lesson timing may not allow young children to finish all the activities on time given their evolving cognitive and motor skills: \textit{``for the young kids, 15-20 minutes is not realistic, because they're not independent. They can't just go off with a partner. They can't read, a lot of them, or it's very beginning reading.''} \commenter{T22} also noted that activities requiring children to make things would take an excessively long time, based on her previous teaching experience. 

Other issues, including digital device requirements and schedule conflicts with other classes, also impacted how teachers felt about the micro-lessons. Specifically, teachers who provided feedback on the finalized micro-lessons were uncertain if some materials would be accessible in schools with heightened regulations on software use and infrastructure. Making resources readily available means working within the current constraints teachers may have on the infrastructure they use for teaching. \commenter{T19}, for instance, noticed that some activities required a Google account. She raised a concern that some schools do not have Google infrastructure: \textit{``I can't remember if anybody can get it or you have to have a Google account, because that could be limiting to some schools and some teachers if kids that are at those schools are not a Google school.''} \commenter{T20} pointed out that some schools had a strict restriction on verifying and using qualified third-party platforms in classrooms, which would possibly impede some micro-lesson activities that use third-party links and resources: \textit{``There are some firewalls, if you will, in terms of schools allowing for outside platforms. It's not that easy just to have accounts for kids, because you're collecting kids' information. Who is the third party? And is this approved? Like [school district], for example, has an approved list of platforms that can be used.''} 

To address the time difficulties, many teachers who implemented the micro-lessons managed class time by cutting or adjusting activities as needed. In every class, \implementer{T15} cut off later activities if the prior ones took a long time, regardless of her willingness to do all of them. \implementer{T18} claimed that she was always \textit{``trying to find time within our schedule''} since the lesson content was compact; she cut off the `Creating Your Digital Citizen Superhero' activity in \textit{Module 1 (Digital Citizenship)} due to time limitations. 

Teachers who provided feedback on the finalized micro-lessons also proposed suggestions for keeping the class at the suggested length. For instance, \commenter{T22} suggested breaking down time-consuming activities, which are supposed to be done in one class, into several pieces and fitting them into several classes: \textit{``If there's any way to break up the posters [activity] over a course of several days, like if they're teaching about critical data literacy, every day you allow them to have some time to add to their poster, that might be a little bit more feasible.''}

Most teachers who implemented the micro-lessons rearranged their class teaching plans to accommodate constraints from teaching teams, schools, and educational sectors. For example, three teachers (\implementer{T6}, \implementer{T17}, \implementer{T18}) merged three sections of content over one week, which was supposed to be taught in three days, into a single, combinatory lesson. \implementer{T6} attributed this change to \textit{``my own scheduling and what grades are doing and how we're supposed to teach and since I'm subbing''}. Teachers who provided feedback on the finalized micro-lessons also brainstormed solutions to external restrictions. \commenter{T21}, when imagining herself implementing the lessons, indicated that she would \textit{``condense the lessons into two weeks''} to cover important topics, considering the class timing requirement from her county.

\subsection{Helping Children Engage With Digital Privacy and Security Concepts Through Micro-Lessons: Advantages, Challenges, and Tailored Teaching Strategies}
\label{subsec:DFteaching}

All teachers we spoke to agreed that micro-lessons could offer children an immersive and accessible learning experience on digital privacy and security concepts. They also identified several challenges from both teachers and children in implementing micro-lessons.

\subsubsection{Micro-Lessons Provide Children Engaging and Memorable Learning Experiences}
\label{subsubsec:DFteachingAdvantage}

Many teachers (iteration 1: 7/7; iteration 2: 3/6) agreed that using micro-lessons was a good approach to improving children's privacy, security, and technology literacy. Most teachers thought the lessons were attractive and engaging with the online videos, pictures, and linked resources on privacy and security topics that were relatable to children's personal experiences.  

Most of the teachers who implemented the lessons reported that the children in their classes enjoyed the micro-lessons and that they were \textit{``really fun for the kids''} (\implementer{T18}). \implementer{T15} said that even children who \textit{``didn't normally have conversations in some of these lessons''} performed actively in the class --- they were \textit{``participating in the conversation and wanted to share.''} This was especially true for the `Digital Citizen Superhero' activities in \textit{Module 1 (Digital Citizenship)}. 
\implementer{T9} enjoyed conducting the `Would You Rather' game in \textit{Module 3 (Digital Privacy)} to engage children in thinking about privacy in their lives. Children in her class participated actively in that discussion, and she found they even extended it after class: \textit{``Even in lunch duty, I hear their conversation and it's like, `Do you want to do this or do you want to do that?' So I think that would you rather is the right item for them.''} \implementer{T5} also shared a moment when the children enjoyed the `Would you Rather' game. She stated that the children had been excited even before she started the instruction. \textit{``Just seeing the excitement and they didn't even know what we were going to do. They just knew they saw `Would You Rather?' They didn't know what the questions were. They didn't know anything. But just to see them excited to have that, to play that.''} 

Moreover, teachers appreciated that the micro-lessons effectively built a connection between what children learned and their experience in daily life with privacy and security issues, attributing it as an important reason for students' high engagement. As \implementer{T6} shared her overall experience of teaching: \textit{``I think the kids were really interested in it and really engaged because it's something that they are used to in everyday life. So they were asking questions.''} \implementer{T17} shared a similar example from her class: \textit{``They actually were very motivated by this entire content, because I think it's very relatable to their real life. I think the whole series, they all have experiences with all of it.''}. Specifically, connecting the activities and making privacy and security concepts relatable to children helped them see how they might encounter issues in their own lives, which could trigger deeper reflections during class. \implementer{T18} shared an example about how children actively related cyberbullying in game playing to what they had personally experienced during the `Roblox Game Chat Simulation' in \textit{Module 1 (Digital Citizenship)}: \textit{``I remember showing them the Roblox and they're like, `Oh I've seen that before.' And they had some good connections so it was easier to relate and how to not cyberbully people.''} 

Teachers who provided feedback on the finalized micro-lessons also speculated activities in micro-lessons could encourage every child, even those who were less proactive in typical classes, to participate in discussions. (\commenter{T19}) contributed, \textit{``The videos are great because most kids... you have the range from really shy, quiet kids to the kids who are always talking and sharing in class. So I think the videos usually hook those kids no matter where they're at in that spectrum.''}

\subsubsection{Making Adjustments for More Comfortable Teaching}
\label{subsubsec:DFteachingLimitTeacher}

Teachers who implemented the micro-lessons (5/7) told us that they sometimes got stuck in the teaching process since they were not familiar with helping children learn about privacy, security, and technology. On one hand, some teachers lacked privacy and security knowledge, which could create discomfort in teaching children about these topics. Because of a limited understanding of related topics, \implementer{T16} admitted that she always doubted if she did the right thing when implementing the class: \textit{``I'm like, did I teach that the right way with the privacy and security? Did I stay on track or did I veer too much from it?...  I don't think I would've felt comfortable necessarily, or knowledgeable about these topics.''} \implementer{T6}, \implementer{T9}, and \implementer{T18} highlighted their limited knowledge of critical data literacy, the focus of \textit{Module 4}. This lack of knowledge made it difficult for them to follow the material, thereby hindering class preparation and implementation. On the other hand, several teachers found themselves unfamiliar with how children use technology. Without enough contextual background, it took these teachers a longer time to prepare for class and to deal with children's reactions during class. For example, \implementer{T16} never played Roblox games before. Since the game was covered in multiple lessons, she researched it on her own but had a difficult time looking up related information in class preparation. 

To ensure they felt comfortable implementing the micro-lessons with their own knowledge about privacy and security in mind, a few teachers who implemented the micro-lessons modified the syllabus to focus on the content they felt comfortable teaching. For example, \implementer{T6} cut off the last lesson of \textit{Module 4 (Critical Data Literacy)} because of her own uncertainty about the topic covered in that lesson: \textit{``I haven't done that last lesson and that's partially because of testing and partially because that's my least comfort.’’}

\subsubsection{Addressing Students' Learning Difficulties for a Better Learning Experience}
\label{subsubsec:DFteachingLimitStudent}

Students sometimes faced challenges in learning privacy and security, which teachers usually found hard to break through. The teachers we spoke to (iteration 1: 7/7; iteration 2: 4/6) reported challenges in students' learning process spanning limited understanding, reading, writing, and comprehension ability, as well as children's limited background knowledge and low awareness of privacy and security risks, which echoed what teachers reported in our formative study (Section \ref{sec:formativefindings}) as well as findings from prior studies (e.g., \cite{kumar2017no, agesilaou2022whose,livingstone2020data}). Regarding children's learning difficulties, they actively sought alternatives and supplements for some of the lesson activities to ensure their students had an effective learning experience.

Perceiving children's growing learning ability during implementation, teachers emphasized the importance of carefully considering if some activities were too advanced for younger children in future lesson development. For instance, \implementer{T6} found it hard to describe the concept of advertisements covered in \textit{Module 2 (Digital Security)} to children: \textit{``I was trying to again relate it to finding ads, `They're trying to sell you things.' But I think again, that's not something that [children] are used to.''} Similarly, \implementer{T9} noted that some topics and activities seem more geared toward older children, so she would turn to Common Sense Media resources to find ways to talk to her younger students about these topics, or think about ways to adjust existing activities, such as having students draw or write out 1-2 sentences or have a class discussion. 
Although we tailored activities to children of different ages in finalized micro-lessons, many teachers felt the settings could be further adjusted for more customization and accessibility, especially for children with special backgrounds, such as English as a second language (\implementer{T5}) or coming from low Internet-access households (\commenter{T23}).

Teachers also highlighted young children's limited focus in the classroom, regarding it as a severe challenge in teaching micro-lessons. Children can easily get distracted from the original goal of privacy and security learning, for which teachers had to make huge efforts to draw back their attention. As \implementer{T18} shared her experience when conducting `Roblox Game Chat Simulation' in the second lesson of \textit{Module 1 (Digital Citizenship)}: \textit{``I know when we were talking about cyberbullying and the Roblox, they were like, `Oh yeah, when I was on Roblox...' Whatever they kept mentioning, I tried to relate it back to privacy and even cyberbullying and what to do.''} Similarly, \commenter{T22} wondered about the `Tower of Treasure' game in \textit{Module 2 (Digital Security)}, claiming that children's learning process might be distracted by gamification: \textit{``a lot of my students would just hit fast-forward on the words and not bother to read it and just focus on collecting the little objects.''} 


Considering children's limited background knowledge, many teachers tended to incorporate alternatives to promote children’s learning from the lessons. During class implementations, several teachers conducted additional activities to provide their students with more context about privacy and security concepts. 
For instance, \implementer{T5} noticed that her students did not understand the concept of bullying, so she conducted an additional small-group activity before introducing cyberbullying, where she and another teacher showed children some facial expression pictures and asked them to describe whether the expressions indicated a nice or a mean attitude. After \implementer{T15} found her students had no basic knowledge about the Internet, she \textit{``extended that conversation a little bit more, talking about how the Internet, [...] `Do you all know what W-W-W stands for?' [...] `We're talking about digital citizenship.'''} Regarding children's knowledge level in their class, teachers who reviewed the finalized micro-lessons also suggested a supplementary warm-up to provide children with more contextual grounding. As \commenter{T23} suggests, \textit{``A kindergartner or first-grader who's just only been on the earth for five or six years, they're like, I know what Google is. But do you know that Google is a search engine?  [...] So adding all of those things are I think, really good modifications to the lessons.''} Regarding the grade bands set in the finalized micro-lessons, \commenter{T22} further suggested that the suggested grade bands could be shifted down for students with special needs.


Noticing that some activities may be complex for young children to conduct, several teachers changed the activities during implementation to make them executable for every child. \implementer{T15}, simplified an activity in \textit{Module 3 (Digital Privacy)}:\textit{``We also did the activity [...] the safe and unsafe behaviors, it was the activity where they had to cut and glue. We didn't cut and glue that day, but what we did, [was] we numbered the responses down at the bottom, and then they attached a number to a safe behavior or an unsafe behavior.''}  Teachers who provided feedback on the finalized micro-lessons also advised modifying activities to be more accessible. \commenter{T24} suggested that \textit{``some of the activities for young kids could be more foundational.''} \commenter{T19} shared the alternative exercise she might do for the `Goal-Setting Worksheet' activity in \textit{Module 2 (Digital Security)}, in which children needed to hand-write responses: \textit{``the little would have a really hard time writing... So, for example, I would take that and put that into Seesaw so the kids could record their answers in Seesaw.''}

In response to children's difficulties with concentration, several teachers applied additions and alternatives when implementing the micro-lessons to ensure their students understood what they learned. For example, Both \implementer{T5} and \implementer{T15} created assessments and reflections that were different from the exit ticket provided in the initial micro-lesson framework, aiming to evaluate how children comprehended the knowledge in every section. \implementer{T17} substituted the `Roblox Game Chat Simulation' in \textit{Module 1 (Digital Citizenship)} with a discussion on cyberbullying on Roblox, fearing the game simulation might distract children from learning goals. Some teachers even conducted extra sessions for revisiting micro-lesson content. \implementer{T16}, for example, conducted the `Drawing Own Digital Citizen Superhero' activity in \textit{Module 1 (Digital Citizenship)} a second time after noticing her students lost focus the first time. She reported that the second implementation was more meaningful than the first: \textit{``The first time it was just like, `Oh, he's a superhero, yay!' Whereas the next time, we stopped and talked about like, `Oh, there's the iPad. What's wrong with him spinning it? What could happen?' 'Well, it's nice that they liked it, but what's going on here?' We stopped and took more time.''} The teachers who provided feedback on the finalized micro-lessons also suggested additional activities to help children consolidate what they learned. For instance, \commenter{T22} indicated that she \textit{``would probably want to have a closure activity or some sort of closure conversation''} after some activities, such as gamification, because the aim of those activities \textit{``were a little bit too abstract''} for children.

\subsection{Micro-Lessons Could Improve Privacy and Security Awareness for Children and Teachers}
\label{subsec:DFtakeaway}

Teachers reported that the micro-lessons did or could enhance children's awareness of digital privacy and security in everyday life, while also acknowledging that their own privacy and security knowledge improved from implementing or reviewing the micro-lessons. 

\subsubsection{
Children's Awareness Around Privacy and Security Improves After Micro-Lessons}
\label{subsubsec:DFtakeawayChildren}

All of the teachers we spoke to acknowledged the potential of the micro-lessons in promoting children's privacy and security literacy. Specifically, teachers saw the potential of micro-lessons providing children with a greater awareness of risks in online activities and more understanding of digital privacy and security in different real-world contexts. They attributed this potential to context-relatable activities and differentiated learning content tailored to different grade levels.

The micro-lessons did cause some shifts in children's thinking and actions around privacy and security in various contexts, as evidenced by the teachers who implemented them. For instance, \implementer{T18} shared, \textit{``It was a good way, when I had those scenarios for them to understand they might think it's okay but it's not. And then they were talking about what's important, why it's important not to share and they're like, `Yeah, you shouldn't talk about your passwords.' ''} \implementer{T5} said her students developed an awareness to question the credibility of information, even if it appeared to be professional: \textit{``...during indoor recess, they (children) were watching a PBS show and it popped up a commercial beforehand for the shampoo. They were like, `Yeah, I don't think just because it's a commercial, I have to believe everything I see...' They definitely have that down pat, they're calling people out on everything.''} 
Teachers who reviewed the finalized micro-lessons speculated similar thinking and behavior changes in children who took the micro-lessons. For example, children might think more before doing every action online after taking the micro-lessons:  \textit{``The main takeaway would be having kids think about, stop and think before you post anything, before you do anything’’} (\commenter{T21}).

Notably, one of the major challenges in helping young children learn about digital privacy and security is to get them to understand how these concepts manifest in everyday technology use, especially considering that some of the concepts are abstract to them. Many teachers felt that the micro-lessons connected privacy and security to children's everyday lives very well, allowing them to get a deeper understanding of related concepts. For example, \implementer{T17} shared a moment in her class when children distinguished confusing concepts related to personal privacy through metaphors: \textit{``They really did understand the concept of personal being something that might be about you but other people share that with you, versus private is something that's very specific to you like your full name, your address. Personal is, oh, I like pizza, but 14 people in my class like pizza.'' }

Moreover, teachers who reviewed the finalized micro-lessons highlighted the potential of content broken out by grade bands, which could offer constant awareness promotion starting from a young age. \commenter{T19} especially appreciated the idea of offering privacy and security lessons to young children. She saw the potential of starting to build this foundational knowledge around digital privacy and security in the earlier grades 
and building on that knowledge over time as technology use evolves to ultimately foster a comprehensive understanding of how to navigate and understand their own expectations and needs for privacy and security as they grow up. She said, \textit{``It might just get an introduction in third grade, but by the time it's in fifth grade, it's kind of really ingrained in them. And so I think there's a large capacity for them to really get a full understanding of all of these things.''} \commenter{T23} shared a similar attitude toward early learning on digital privacy and security: \textit{``It would be a benefit to use those types of lessons in a school to be able to just start the conversation and start the understanding for our younger learners of what it means to be a good technology scholar in the school and how we can use it.''}


\subsubsection{Teachers' Privacy and Security Awareness Improves After Micro-Lessons}
\label{subsubsec:DFtakeawayTeacher}

Aside from improving children's understanding and practice of privacy, security, and digital literacy, many teachers we spoke to (iteration 1: 6/7; iteration 2: 3/6) also reported that they themselves could benefit from implementing or reviewing the micro-lessons. Their reported takeaways span two aspects: getting to know about children's current privacy and security literacy and practice, and learning about digital privacy and security concepts themselves.

Teachers believed that implementing micro-lessons could help them learn about children’s current experiences as digital citizens and attitudes toward online risks, as well as determine their awareness and grasp of digital privacy and security-related topics. \commenter{T21} described how implementing the lessons might help her understand more about her students' knowledge and experiences using technology: ``\textit{I think I would learn a little bit more of what [my students] already know about the online world, and maybe what their experiences are. What they don't know, your digital footprint and stuff, how many of them are aware of that? I think that's what I would find out and that I'd be really curious about that aspect.}'' Many teachers agreed that getting this information would be very useful for them to navigate teaching privacy and security concepts. Teachers who implemented the micro-lessons shared some in-class moments when they learned about children's experiences. \implementer{T6} mentioned that the `Would You Rather' game in \textit{Module 3 (Digital Privacy)} helped her better gauge students’ thoughts around digital privacy and security through discussions. \implementer{T5} found the children’s personal stories about experiencing and solving online risks such as cyberbullying eye-opening: \textit{``I could see them as not just being kids but realizing that they really are trying to figure things out.''} Additionally, some of the teachers regarded unpacking children's privacy and security experiences as a way to reflect on children's learning needs. When \implementer{T15} talked about her general takeaways from teaching her class, she shared her shock that the children have far less awareness of threats from online strangers than she thought: \textit{``I was surprised by some things that they didn't know...could be potentially dangerous. Because you're playing with someone online who may be far away they're thinking, `Well, if they're far away, what's the big deal?'''}, for which she argued that \textit{``there has to be more discussion, even more now than before because they're online so much.''} 

Some teachers felt that even without implementing micro-lessons, the instructional document could supplement their own professional development around digital privacy and security. In \commenter{T19}'s words: \textit{``I think teachers who don't know much about [digital privacy and security] could actually watch these and learn a lot about them. So I think there's potential for a teacher with no idea about this stuff to do the lessons first themselves and get a lot of `aha' moments.''} Further, several teachers noted that implementing micro-lessons with their students helped them deepen their privacy and security knowledge. \implementer{T9} said that in doing the micro-lessons, \textit{``I learned that I have no idea what critical data literacy is.’’} When guiding children playing the `Roblox Game Chat Simulation' in \textit{Module 1 (Digital Citizenship)}, \implementer{T16} discovered that cyberbullying was more prevalent in children's online gameplay than she assumed. Similarly, \implementer{T17} indicated that the micro-lessons triggered a deeper understanding of online safety: \textit{``I definitely think it made me think a lot about this type of stuff that I probably haven't thought about as an adult. I think things you think about for a kid... I don't think about my safety as much online as maybe I should be.'' } 
When \commenter{T22} reviewed the finalized micro-lessons, she recalled how children pointed out her oversights in a privacy lesson, which she regarded as a possible learning moment for other teachers if had implemented the micro-lessons: \textit{``Every time I teach a lesson like this, my students will always call me out on the things that I do that I shouldn't according to the lesson, like using the same six passwords across many websites.''}

\section{Discussion}



We present the design and evaluation of digital privacy and security micro-lessons for elementary school children, and provide evidence from teachers on the potential of integrating micro-lessons to support in-classroom privacy and security education. This evidence supports prior works advocating for contextual-based educational resources~\cite{blinder2024evaluating, alghythee2024towards, kumar2019privacy, kumar2020strengthening} and classroom-based learning~\cite{hartikainen2019children,lastdrager2017effective,livingstone2020data,rahman2020importance} in privacy and security education for children. Below, we provide grounded design implications for future privacy and security education in elementary schools. 


\subsection{Breaking Out Privacy, Security, and Critical Data Literacy Content From Digital Citizenship Curriculum}
\label{subsec:discusspsdl}
Our findings suggest that micro-lessons designed to encourage children to continually reflect on and examine their digital privacy and security experiences can be beneficial in helping them become familiar with these concepts over time. This early introduction, especially before they encounter more significant external privacy and security threats in their teen years, can be particularly valuable (Section \ref{subsubsec:DFtakeawayChildren}). 

Moving beyond prior suggestions of incorporating privacy and security learning in elementary school classrooms~\cite{hartikainen2019children,lastdrager2017effective,livingstone2020data,rahman2020importance}, we demonstrate the utility of implementing digital privacy and security specialized micro-lessons. While acknowledging existing digital literacy curriculum (e.g., via Common Sense Media~\cite{commonsense_digital_citizenship}), we also note that these lessons typically focus more on digital citizenship and training children to be aware of potential online risks. While this is certainly beneficial, the teachers in our study appreciated the deeper dives into privacy and security topics (Section \ref{sec:formative}). In particular, with the rise of Artificial Intelligence (AI), children need to learn critical data literacy skills which refers to the ability to read, interpret, critique, and make informed decisions based on data~\cite{casal2023ai}. 

Owing to our findings and these trends, we suggest that digital privacy, security, and critical data literacy need to be broken out as independent learning topics for K--8 children. Gaining foundational knowledge on these topics will help children develop strong digital literacy and digital citizenship skills and enable them to navigate an evolving technological landscape as they become teenagers.

\subsection{Including Children's Relatable Everyday Life in Learning Process}
During the evaluation studies, teachers identified the benefits of including life-relevant examples in micro-lessons to help children engage in digital privacy and security learning in elementary school classes. For instance, teachers appreciated the privacy-oriented `Would You Rather' scenarios and cyberbullying examples situated into the Roblox emulator. Teachers reflected on how these life-relevant examples helped children to more deeply discuss and relate to privacy and security as it manifests in their own day-to-day experiences online (Section \ref{subsubsec:DFteachingAdvantage}). Our findings echo evidence from prior CCI research that context-based approaches can facilitate children's privacy and security learning~\cite{blinder2024evaluating, alghythee2024towards}. Education researchers have already recognized that context-based learning --- via mock-up or real context --- fosters K--12 students' ability to discover, analyze, and solve problems better than traditional learning in academic STEM subjects~\cite{sevian2018does, yu2015enhancing}. Since privacy and security are often correlated to specific social contexts~\cite{Nissenbaum_2010}, having children discuss different situations in which they encounter privacy and security issues can help them develop a sense of their boundaries~\cite{blinder2024evaluating}. Rather than teaching conceptual privacy and security knowledge only, we argue that exposing children to discussions and activities focused more on norms of dealing with privacy and security risks in different scenarios can aid learning over time~\cite{kumar2020strengthening}. 



We suggest two ways for learning material developers and educators to integrate appropriate and timely examples into in-classroom privacy and security learning. First, aligning with findings from Kumar et al.~\cite{kumar2023understanding}, children should be involved in the development of any learning activities and interactive learning tools. We note that some of the micro-lesson components we reference were previously co-designed with children \cite{blinder2024evaluating}. Moreover, our findings suggest students' needs vary from classroom to classroom, especially for children in different grades or with different backgrounds (Section \ref{subsec:need2}, Section \ref{subsubsec:DFteachingLimitStudent}). Educators should identify children's learning needs and emerging scenarios by having explicit and periodical conversations with students about their technology experiences, starting when they are young. Second, with the development of technology-integrated classrooms, there are emerging privacy and security concerns in classroom activities resulting from both children's behavior (e.g., forgetting to log out of school devices)~\cite{kumar2017no, kumar2019privacy, bacak2022elementary}, and the breaches in sociotechnical systems (e.g., surveillance from educational software)~\cite{chanenson2023uncovering, lu2021coding}. If teachers and schools are open to it, many of the privacy and security incidents encountered in the classroom could be directly integrated into ad-hoc teachable moments on these topics. 


As with many interactive learning tools, there are trade-offs in introducing privacy and security concepts using approaches that are not too abstract and are easily relatable to children's own experiences. Our teacher participants described how children could become easily distracted from the original learning goals, such as with the `Roblox Game Chat Simulation' (Section \ref{subsubsec:DFteachingLimitStudent}). Therefore, we suggest both learning material developers and educators pay extra attention to managing the distractions that vivid learning materials may bring to children. When designing and building activities, games, and tools for privacy and security education, developers should balance the trade-off between playfulness and learning effectiveness. For example, interactive games could have attractive playing styles and visual elements for an immersive learning experience but should not be complex and hard to relate back to privacy and security concepts. Moreover, micro-lessons could also include concrete examples for teachers of how to bring children's attention back to digital privacy and security during class if they have gotten off focus. For teachers, additional support may be needed in thinking through the preparation for and timing of their lessons, as well as in reorienting children's focus on the lesson concepts. Examples of how to handle these challenges could include having time to discuss or be excited about the everyday connection before delving deeper into the lessons; implementing the micro-lessons at the end of the day; or using questions during activities to refocus on learning goals.

\subsection{Conducting Additional Professional Development for Teachers} 
Teachers who evaluated our micro-lessons reported benefits from the detailed instructions and concrete teacher resources, which not only facilitated their teaching preparation (Section \ref{subsubsec:DFinfrastructureFacilitator}), but also enriched their knowledge of digital privacy and security (Section \ref{subsubsec:DFtakeawayTeacher}). However, they also expressed discomfort with teaching due to their limited knowledge of certain topics (Section \ref{subsubsec:DFteachingLimitTeacher}), aligning with prior reports on teacher's limited understanding of digital privacy and security concept when educating children~\cite{kumar2019privacy}. Meanwhile, prior works also found teachers tend to focus on the safety side rather than the technical side when teaching digital privacy and security~\cite{smith2023educators, mcleod2024comparing}, implying their limited knowledge of the technical mechanism of privacy and security. 

Therefore, teaching guidance and teacher resources are not enough to ensure a smooth digital privacy and security learning experience in the classroom. To equip teachers with sufficient knowledge to help children learn about digital privacy and security, they also require frequent professional development on these topics, with regular updates to account for the ever-changing nature of technology. Moreover, in our evaluation, teachers were especially unfamiliar with critical data literacy, even though it is an important skill in digital practice and should be included in digital privacy and security education (Section \ref{subsec:discusspsdl}). We, therefore, suggest a focus on critical data literacy in professional development. 

\subsection{Tailoring Teaching Schedules to Overcome Infrastructural Restrictions}
In our evaluation studies, teachers spoke of how in elementary schools, teacher teams usually take responsibility for developing and executing teaching activities, following the policies formed by the school or external education sectors. When integrating digital privacy and security education into elementary school classrooms, the teaching approaches must comply with existing policies. Therefore, teachers implementing and providing feedback on the micro-lessons raised concerns about implementing curriculum under existing constraints, such as those on a strict class schedule and digital device usage (Section \ref{subsubsec:DFinfrastructureDifficulty}). 

Specifically, the requirements for digital infrastructure could be a significant blocker of implementing a universal digital privacy and security curriculum. In the US, several educational technology providers, including Google, Microsoft, and Apple, collaborate with elementary schools to create digital classrooms. Until now, there are no national standards that designate devices, systems, and tools used~\cite{EdWeek_K12_Market}. Due to the funding constraints of schools and overlapping functions of products, many schools do not collaborate with all providers. 
Moreover, regulations aimed at protecting student safety, such as the US Family Educational Rights and Privacy Act (FERPA)~\cite{ED_FERPA}, have necessitated rigorous ethical and legal considerations when integrating new digital tools into elementary school classrooms~\cite{atabey2024international}. Given that many privacy and security educational materials and tools are digital and provided by various entities, some activities of a universal curriculum may be restricted within a classroom due to digital infrastructure limitations. 

Instead of developing a complicated lesson plan that fits every school's infrastructural requirements, which could be costly, we suggest teacher teams actively tailor the lesson schedule and content to suit the special needs of their schools. This could include reorganizing the structure or adjusting the duration of lessons, and seeking or creating alternatives for the restricted activities, as teachers in our study did when implementing the micro-lessons (Section \ref{subsubsec:DFinfrastructureDifficulty}). We note that adjusting class schedules and activities may make the micro-lessons diverge from their original teaching model and learning objectives, compromising children's learning outcomes eventually. For example, combining all micro-lessons into one large session of a few hours, which is similar to traditional one-off training, would fail to make privacy and security learning a scaffolded  process as we proposed. Therefore, when tailoring lessons, we suggest teaching teams follow the learning objective we provided for each micro-lesson and each learning module, actively discussing within the team and with technology experts if possible, and testing the tailored lessons with small groups of children before classroom-wide teaching. 

Moreover, when developing interactive digital tools for in-classroom privacy and security education, we recommend developers carefully consider the digital infrastructural restrictions at elementary schools. First, the tools should be easy to configure and use based on a fundamental digital system settings at school. Applications that rely on extensive computing resources, complicated configurations, and advanced hardware like virtual reality headsets may offer children a more attractive learning experience but could be hard to deploy within a typical classroom and difficult to apply universally. Second, developers should design and build their any technical learning materials or applications to adhere to school policies and regional regulations on in-school digital service use. For example, any data transition and storage within the applications should not use insecure servers that policies do not allow. If possible, the applications could eliminate unnecessary data collection and processing, such as requiring logins and student demographic information collection, to avoid policy violations. Third, and more importantly, any technical resources could quickly become outdated so any technical artifacts/learning materials should be created with an eye to have a relatively long shelf life and provide means to update the materials over time.





\section{Limitation and Future Work}
Below, we present several limitations of our work, which could be addressed in future work. 

First, our micro-lessons focus on four main topics---digital citizenship, digital security, digital privacy, and critical data literacy. Future lesson developments should be expanded to address related nuanced issues such as misinformation and fake news, privacy and security in algorithms and generative AI, and more. 

While our study enabled us to get a rich sample of data from nearly 40 teachers, additional work in development and evaluation should be done by involving more educators, technology experts, and children to make micro-lessons more robust and usable. For instance, our micro-lesson evaluation only involved seven teachers who implemented lessons in their classes for a short term, with another six teachers reviewing the instructional document and providing feedback without implementing the lessons. Additionally, we did not collect feedback or measure learning outcomes directly from the children's side. Future evaluations could involve more educators who implement the revised lessons and technology experts in designing additional privacy and security-related activities and further differentiating activities tailored towards different grade bands. Future evaluations could also be conducted with a wider range of schools in varied geographical areas and measure learning outcomes more formally. 

\section{Conclusion}
Prior work on children's digital privacy and security has largely focused on understanding children's and teachers' needs~\cite{kumar2019privacy,martin2023teacher, bacak2022elementary, nicholson2020investigating} or designing individual systems or experiences~\cite{kumar2017no, ge2022treat, zhang2017cyberheroes, yap2020phone}. In this paper, we leveraged co-design with teachers to iteratively develop a set of micro-lessons to support children's digital privacy and security learning at school. Through our analysis of the design and evaluation process, we highlight several important findings in the study of privacy and security education for Grades K--8 children. First, we provide support for the potential of short, contextualized micro-lessons integrated flexibly into classrooms to help young children accumulate privacy and security skills and their application over time. We identify a wide range of life-relevant topics that are connected with students' everyday lives, and we encourage teachers to continue exploring connections between their students' interests and experiences and the underlying goals of the micro-lessons. With the growing push for socio-emotional learning (SEL) experiences in elementary schools, as well as the increasing reliance on hardware and software to support learning, we expect the micro-lesson content will only become more important over time.

In addition to benefits for students, we also find ways these micro-lessons can enhance teachers' experiences in the classroom. Our analysis suggests a potential for the micro-lessons to help teachers develop understanding and fluency with privacy and security concepts, as well as develop awareness and familiarity of the everyday privacy and security risks their students face and the understanding of their students. Finally, we stress that 
helping children learn this content also requires more investment in professional development, both to ensure teachers feel comfortable teaching these topics and to identify ways to integrate lessons into a wide variety of schools with different devices and setups. More work is needed to tailor such lessons and activities to teachers' and students' diverse needs in the classroom, to different grade-band levels, and to develop connections to school technology policies. 


\bibliographystyle{ACM-Reference-Format}
\bibliography{Main}


\begin{thebibliography}{90}


\ifx \showCODEN    \undefined \def \showCODEN     #1{\unskip}     \fi
\ifx \showDOI      \undefined \def \showDOI       #1{#1}\fi
\ifx \showISBNx    \undefined \def \showISBNx     #1{\unskip}     \fi
\ifx \showISBNxiii \undefined \def \showISBNxiii  #1{\unskip}     \fi
\ifx \showISSN     \undefined \def \showISSN      #1{\unskip}     \fi
\ifx \showLCCN     \undefined \def \showLCCN      #1{\unskip}     \fi
\ifx \shownote     \undefined \def \shownote      #1{#1}          \fi
\ifx \showarticletitle \undefined \def \showarticletitle #1{#1}   \fi
\ifx \showURL      \undefined \def \showURL       {\relax}        \fi
\providecommand\bibfield[2]{#2}
\providecommand\bibinfo[2]{#2}
\providecommand\natexlab[1]{#1}
\providecommand\showeprint[2][]{arXiv:#2}

\bibitem[Agesilaou and Kyza(2022)]%
        {agesilaou2022whose}
\bibfield{author}{\bibinfo{person}{Andria Agesilaou} {and} \bibinfo{person}{Eleni~A Kyza}.} \bibinfo{year}{2022}\natexlab{}.
\newblock \showarticletitle{Whose data are they? Elementary school students’ conceptualization of data ownership and privacy of personal digital data}.
\newblock \bibinfo{journal}{\emph{International Journal of Child-Computer Interaction}}  \bibinfo{volume}{33} (\bibinfo{year}{2022}).
\newblock
\urldef\tempurl%
\url{https://doi.org/10.1016/j.ijcci.2022.100462}
\showDOI{\tempurl}


\bibitem[Alghythee et~al\mbox{.}(2024)]%
        {alghythee2024towards}
\bibfield{author}{\bibinfo{person}{Kenan Kamel~A Alghythee}, \bibinfo{person}{Adel Hrncic}, \bibinfo{person}{Karthik Singh}, \bibinfo{person}{Sumanth Kunisetty}, \bibinfo{person}{Yaxing Yao}, {and} \bibinfo{person}{Nikita Soni}.} \bibinfo{year}{2024}\natexlab{}.
\newblock \showarticletitle{Towards Understanding Family Privacy and Security Literacy Conversations at Home: Design Implications for Privacy Literacy Interfaces}. In \bibinfo{booktitle}{\emph{Proceedings of the CHI Conference on Human Factors in Computing Systems}} (Honolulu, HI, USA) \emph{(\bibinfo{series}{CHI '24})}. \bibinfo{publisher}{Association for Computing Machinery}, \bibinfo{address}{New York, NY, USA}, Article \bibinfo{articleno}{983}, \bibinfo{numpages}{12}~pages.
\newblock
\showISBNx{9798400703300}
\urldef\tempurl%
\url{https://doi.org/10.1145/3613904.3641962}
\showDOI{\tempurl}


\bibitem[Assal et~al\mbox{.}(2018)]%
        {assal2018exploration}
\bibfield{author}{\bibinfo{person}{Hala Assal}, \bibinfo{person}{Ahsan Imran}, {and} \bibinfo{person}{Sonia Chiasson}.} \bibinfo{year}{2018}\natexlab{}.
\newblock \showarticletitle{An exploration of graphical password authentication for children}.
\newblock \bibinfo{journal}{\emph{International Journal of Child-Computer Interaction}}  \bibinfo{volume}{18} (\bibinfo{date}{Nov.} \bibinfo{year}{2018}), \bibinfo{pages}{37–46}.
\newblock
\showISSN{2212-8689}
\urldef\tempurl%
\url{https://doi.org/10.1016/j.ijcci.2018.06.003}
\showDOI{\tempurl}


\bibitem[Atabey and Hooper(2024)]%
        {atabey2024international}
\bibfield{author}{\bibinfo{person}{Ay{\c{c}}a Atabey} {and} \bibinfo{person}{Louise Hooper}.} \bibinfo{year}{2024}\natexlab{}.
\newblock \bibinfo{booktitle}{\emph{International regulatory decisions concerning EdTech companies’ data practices}}.
\newblock \bibinfo{type}{{T}echnical {R}eport}. \bibinfo{institution}{Digital Futures for Children centre, 5Rights Foundation}.
\newblock
\urldef\tempurl%
\url{http://eprints.lse.ac.uk/123805/1/DFC_Brief_International_regulatory_decisions_final.pdf}
\showURL{%
\tempurl}


\bibitem[Auxier et~al\mbox{.}(2020)]%
        {Auxier_2020}
\bibfield{author}{\bibinfo{person}{Brooke Auxier}, \bibinfo{person}{Monica Anderson}, \bibinfo{person}{Andrew Perrin}, {and} \bibinfo{person}{Erica Turner}.} \bibinfo{year}{2020}\natexlab{}.
\newblock \bibinfo{booktitle}{\emph{Parenting Children in the Age of Screens}}.
\newblock \bibinfo{type}{{T}echnical {R}eport}. \bibinfo{institution}{Pew Research Center}.
\newblock
\urldef\tempurl%
\url{https://www.pewresearch.org/internet/2020/07/28/parenting-children-in-the-age-of-screens/}
\showURL{%
\tempurl}


\bibitem[Bacak et~al\mbox{.}(2022)]%
        {bacak2022elementary}
\bibfield{author}{\bibinfo{person}{Julie Bacak}, \bibinfo{person}{Florence Martin}, \bibinfo{person}{Lynn Ahlgrim-Delzell}, \bibinfo{person}{Drew Polly}, {and} \bibinfo{person}{WeiChao Wang}.} \bibinfo{year}{2022}\natexlab{}.
\newblock \showarticletitle{Elementary educator perceptions of student digital safety based on technology use in the classroom}.
\newblock \bibinfo{journal}{\emph{Computers in the Schools}} \bibinfo{volume}{39}, \bibinfo{number}{2} (\bibinfo{year}{2022}), \bibinfo{pages}{186--202}.
\newblock
\urldef\tempurl%
\url{https://doi.org/10.1080/07380569.2022.2071233}
\showDOI{\tempurl}


\bibitem[Bilstrup et~al\mbox{.}(2022)]%
        {bilstrup2022supporting}
\bibfield{author}{\bibinfo{person}{Karl-Emil~Kj\ae{}r Bilstrup}, \bibinfo{person}{Magnus~H\o{}holt Kaspersen}, \bibinfo{person}{Mille~Skovhus Lunding}, \bibinfo{person}{Marie-Monique Schaper}, \bibinfo{person}{Maarten Van~Mechelen}, \bibinfo{person}{Mariana~Aki Tamashiro}, \bibinfo{person}{Rachel~Charlotte Smith}, \bibinfo{person}{Ole~Sejer Iversen}, {and} \bibinfo{person}{Marianne~Graves Petersen}.} \bibinfo{year}{2022}\natexlab{}.
\newblock \showarticletitle{Supporting Critical Data Literacy in K-9 Education: Three Principles for Enriching Pupils’ Relationship to Data}. In \bibinfo{booktitle}{\emph{Proceedings of the 21st Annual ACM Interaction Design and Children Conference}} (Braga, Portugal) \emph{(\bibinfo{series}{IDC '22})}. \bibinfo{publisher}{Association for Computing Machinery}, \bibinfo{address}{New York, NY, USA}, \bibinfo{pages}{225–236}.
\newblock
\showISBNx{9781450391979}
\urldef\tempurl%
\url{https://doi.org/10.1145/3501712.3530783}
\showDOI{\tempurl}


\bibitem[Blinder et~al\mbox{.}(2024)]%
        {blinder2024evaluating}
\bibfield{author}{\bibinfo{person}{Elana~B. Blinder}, \bibinfo{person}{Marshini Chetty}, \bibinfo{person}{Jessica Vitak}, \bibinfo{person}{Zoe Torok}, \bibinfo{person}{Salina Fessehazion}, \bibinfo{person}{Jason Yip}, \bibinfo{person}{Jerry~Alan Fails}, \bibinfo{person}{Elizabeth Bonsignore}, {and} \bibinfo{person}{Tamara Clegg}.} \bibinfo{year}{2024}\natexlab{}.
\newblock \showarticletitle{Evaluating the Use of Hypothetical 'Would You Rather' Scenarios to Discuss Privacy and Security Concepts with Children}.
\newblock \bibinfo{journal}{\emph{Proc. ACM Hum.-Comput. Interact.}} \bibinfo{volume}{8}, \bibinfo{number}{CSCW1}, Article \bibinfo{articleno}{165} (\bibinfo{date}{apr} \bibinfo{year}{2024}), \bibinfo{numpages}{32}~pages.
\newblock
\urldef\tempurl%
\url{https://doi.org/10.1145/3641004}
\showDOI{\tempurl}


\bibitem[Braun and Clarke(2006)]%
        {braun2006using}
\bibfield{author}{\bibinfo{person}{Virginia Braun} {and} \bibinfo{person}{Victoria Clarke}.} \bibinfo{year}{2006}\natexlab{}.
\newblock \showarticletitle{Using thematic analysis in psychology}.
\newblock \bibinfo{journal}{\emph{Qualitative Research in Psychology}} \bibinfo{volume}{3}, \bibinfo{number}{2} (\bibinfo{year}{2006}), \bibinfo{pages}{77--101}.
\newblock
\urldef\tempurl%
\url{https://doi.org/10.1191/1478088706qp063oa}
\showDOI{\tempurl}


\bibitem[Braun and Clarke(2019)]%
        {braun2019reflecting}
\bibfield{author}{\bibinfo{person}{Virginia Braun} {and} \bibinfo{person}{Victoria Clarke}.} \bibinfo{year}{2019}\natexlab{}.
\newblock \showarticletitle{Reflecting on reflexive thematic analysis}.
\newblock \bibinfo{journal}{\emph{Qualitative Research in Sport, Exercise and Health}} \bibinfo{volume}{11}, \bibinfo{number}{4} (\bibinfo{year}{2019}), \bibinfo{pages}{589--597}.
\newblock
\urldef\tempurl%
\url{https://doi.org/10.1080/2159676X.2019.1628806}
\showDOI{\tempurl}


\bibitem[Bybee et~al\mbox{.}(2006)]%
        {bybee2006bscs}
\bibfield{author}{\bibinfo{person}{Rodger~W Bybee}, \bibinfo{person}{Joseph~A Taylor}, \bibinfo{person}{April Gardner}, \bibinfo{person}{Pamela Van~Scotter}, \bibinfo{person}{J~Carlson Powell}, \bibinfo{person}{Anne Westbrook}, {and} \bibinfo{person}{Nancy Landes}.} \bibinfo{year}{2006}\natexlab{}.
\newblock \bibinfo{booktitle}{\emph{The BSCS 5E instructional model: Origins and effectiveness}}.
\newblock \bibinfo{type}{{T}echnical {R}eport}. \bibinfo{institution}{Office of Science Education National Institutes of Health}.
\newblock
\urldef\tempurl%
\url{https://bscs.org/reports/the-bscs-5e-instructional-model-origins-and-effectiveness/}
\showURL{%
\tempurl}


\bibitem[Casal-Otero et~al\mbox{.}(2023)]%
        {casal2023ai}
\bibfield{author}{\bibinfo{person}{Lorena Casal-Otero}, \bibinfo{person}{Alejandro Catala}, \bibinfo{person}{Carmen Fern{\'a}ndez-Morante}, \bibinfo{person}{Maria Taboada}, \bibinfo{person}{Beatriz Cebreiro}, {and} \bibinfo{person}{Sen{\'e}n Barro}.} \bibinfo{year}{2023}\natexlab{}.
\newblock \showarticletitle{AI literacy in K-12: a systematic literature review}.
\newblock \bibinfo{journal}{\emph{International Journal of STEM Education}} \bibinfo{volume}{10}, \bibinfo{number}{1} (\bibinfo{year}{2023}), \bibinfo{pages}{29}.
\newblock
\urldef\tempurl%
\url{https://doi.org/10.1186/s40594-023-00418-7}
\showDOI{\tempurl}


\bibitem[Cavanagh(2017)]%
        {EdWeek_K12_Market}
\bibfield{author}{\bibinfo{person}{Sean Cavanagh}.} \bibinfo{year}{2017}\natexlab{}.
\newblock \showarticletitle{Amazon, Apple, Google, and Microsoft Battle for K-12 Market, and Loyalties of Educators}.
\newblock \bibinfo{journal}{\emph{EdWeek Market Brief}} (\bibinfo{date}{May} \bibinfo{year}{2017}).
\newblock
\urldef\tempurl%
\url{https://marketbrief.edweek.org/sales-marketing/amazon-apple-google-and-microsoft-battle-for-k-12-market-and-loyalties-of-educators/2017/05}
\showURL{%
\tempurl}


\bibitem[Chanenson et~al\mbox{.}(2023)]%
        {chanenson2023uncovering}
\bibfield{author}{\bibinfo{person}{Jake Chanenson}, \bibinfo{person}{Brandon Sloane}, \bibinfo{person}{Navaneeth Rajan}, \bibinfo{person}{Amy Morril}, \bibinfo{person}{Jason Chee}, \bibinfo{person}{Danny~Yuxing Huang}, {and} \bibinfo{person}{Marshini Chetty}.} \bibinfo{year}{2023}\natexlab{}.
\newblock \showarticletitle{Uncovering Privacy and Security Challenges In K-12 Schools}. In \bibinfo{booktitle}{\emph{Proceedings of the 2023 CHI Conference on Human Factors in Computing Systems}} (Hamburg, Germany) \emph{(\bibinfo{series}{CHI '23})}. \bibinfo{publisher}{Association for Computing Machinery}, \bibinfo{address}{New York, NY, USA}, Article \bibinfo{articleno}{592}, \bibinfo{numpages}{28}~pages.
\newblock
\showISBNx{9781450394215}
\urldef\tempurl%
\url{https://doi.org/10.1145/3544548.3580777}
\showDOI{\tempurl}


\bibitem[Cino and Vandini(2020)]%
        {cino2020does}
\bibfield{author}{\bibinfo{person}{Davide Cino} {and} \bibinfo{person}{Chiara~Dalledonne Vandini}.} \bibinfo{year}{2020}\natexlab{}.
\newblock \showarticletitle{“Why Does a Teacher Feel the Need to Post My Kid?”: Parents and Teachers Constructing Morally Acceptable Boundaries of Children’s Social Media Presence}.
\newblock \bibinfo{journal}{\emph{International Journal of Communication}} \bibinfo{volume}{14}, \bibinfo{number}{00} (\bibinfo{date}{Feb.} \bibinfo{year}{2020}), \bibinfo{pages}{1153--1172}.
\newblock
\urldef\tempurl%
\url{https://ijoc.org/index.php/ijoc/article/view/12493}
\showURL{%
\tempurl}


\bibitem[Culnan and Carlin(2009)]%
        {culnan2009online}
\bibfield{author}{\bibinfo{person}{Mary~J. Culnan} {and} \bibinfo{person}{Thomas~J. Carlin}.} \bibinfo{year}{2009}\natexlab{}.
\newblock \showarticletitle{Online privacy practices in higher education: making the grade?}
\newblock \bibinfo{journal}{\emph{Commun. ACM}} \bibinfo{volume}{52}, \bibinfo{number}{3} (\bibinfo{date}{Mar} \bibinfo{year}{2009}), \bibinfo{pages}{126–130}.
\newblock
\showISSN{0001-0782}
\urldef\tempurl%
\url{https://doi.org/10.1145/1467247.1467277}
\showDOI{\tempurl}


\bibitem[Desimpelaere et~al\mbox{.}(2020)]%
        {desimpelaere2020knowledge}
\bibfield{author}{\bibinfo{person}{Laurien Desimpelaere}, \bibinfo{person}{Liselot Hudders}, {and} \bibinfo{person}{Dieneke Van~de Sompel}.} \bibinfo{year}{2020}\natexlab{}.
\newblock \showarticletitle{Knowledge as a strategy for privacy protection: How a privacy literacy training affects children's online disclosure behavior}.
\newblock \bibinfo{journal}{\emph{Computers in Human Behavior}}  \bibinfo{volume}{110} (\bibinfo{year}{2020}), \bibinfo{pages}{106382}.
\newblock
\urldef\tempurl%
\url{https://doi.org/10.1016/j.chb.2020.106382}
\showDOI{\tempurl}


\bibitem[Drader(2022)]%
        {drader2022digital}
\bibfield{author}{\bibinfo{person}{Sherry~L Drader}.} \bibinfo{year}{2022}\natexlab{}.
\newblock \bibinfo{booktitle}{\emph{Digital Citizenship for Elementary Students}}.
\newblock \bibinfo{type}{{T}echnical {R}eport}. \bibinfo{institution}{Educational Leadership Student}.
\newblock
\urldef\tempurl%
\url{https://cedar.wwu.edu/edlead_stuschol/1}
\showURL{%
\tempurl}


\bibitem[Education(2024)]%
        {commonsense_digital_citizenship}
\bibfield{author}{\bibinfo{person}{Common~Sense Education}.} \bibinfo{year}{2024}\natexlab{}.
\newblock \bibinfo{title}{Digital Citizenship}.
\newblock
\newblock
\urldef\tempurl%
\url{https://www.commonsense.org/education/digital-citizenship}
\showURL{%
\tempurl}
\newblock
\shownote{Accessed: 2024-05}.


\bibitem[for Education~Statistics(2023)]%
        {nces2023publicteachers}
\bibfield{author}{\bibinfo{person}{National~Center for Education~Statistics}.} \bibinfo{year}{2023}\natexlab{}.
\newblock \bibinfo{title}{Characteristics of Public School Teachers}.
\newblock
\newblock
\urldef\tempurl%
\url{https://nces.ed.gov/programs/coe/indicator/clr/public-school-teachers}
\showURL{%
\tempurl}
\newblock
\shownote{Accessed: 2024-06}.


\bibitem[for Education~Statistics(2024)]%
        {Montpelier}
\bibfield{author}{\bibinfo{person}{National~Center for Education~Statistics}.} \bibinfo{year}{2024}\natexlab{}.
\newblock \bibinfo{title}{School Profiles}.
\newblock
\newblock
\urldef\tempurl%
\url{https://nces.ed.gov/ccd/schoolsearch/}
\showURL{%
\tempurl}
\newblock
\shownote{Accessed: 2024-07-01}.


\bibitem[Goldman et~al\mbox{.}(2022)]%
        {goldman2022collaborative}
\bibfield{author}{\bibinfo{person}{Susan~R Goldman}, \bibinfo{person}{Cindy~E Hmelo-Silver}, {and} \bibinfo{person}{Eleni~A Kyza}.} \bibinfo{year}{2022}\natexlab{}.
\newblock \showarticletitle{Collaborative Design as a context for teacher and researcher learning: introduction to the special issue}.
\newblock \bibinfo{journal}{\emph{Cognition and Instruction}} \bibinfo{volume}{40}, \bibinfo{number}{1} (\bibinfo{year}{2022}), \bibinfo{pages}{1--6}.
\newblock
\urldef\tempurl%
\url{https://doi.org/10.1080/07370008.2021.2010215}
\showDOI{\tempurl}


\bibitem[{Google}(2024)]%
        {google_be_internet_awesome}
\bibfield{author}{\bibinfo{person}{{Google}}.} \bibinfo{year}{2024}\natexlab{}.
\newblock \bibinfo{title}{Be Internet Awesome - A Program to Teach Kids Online Safety}.
\newblock
\newblock
\urldef\tempurl%
\url{https://beinternetawesome.withgoogle.com/en_us}
\showURL{%
\tempurl}
\newblock
\shownote{Accessed: 2024-05}.


\bibitem[Hartikainen et~al\mbox{.}(2019)]%
        {hartikainen2019children}
\bibfield{author}{\bibinfo{person}{Heidi Hartikainen}, \bibinfo{person}{Netta Iivari}, {and} \bibinfo{person}{Marianne Kinnula}.} \bibinfo{year}{2019}\natexlab{}.
\newblock \showarticletitle{Children’s design recommendations for online safety education}.
\newblock \bibinfo{journal}{\emph{International Journal of Child-Computer Interaction}}  \bibinfo{volume}{22} (\bibinfo{year}{2019}), \bibinfo{pages}{100146}.
\newblock
\urldef\tempurl%
\url{https://doi.org/10.1016/j.ijcci.2019.100146}
\showDOI{\tempurl}


\bibitem[Hasebrink et~al\mbox{.}(2009)]%
        {hasebrink2009comparing}
\bibfield{author}{\bibinfo{person}{Uwe Hasebrink}, \bibinfo{person}{Sonia Livingstone}, \bibinfo{person}{Leslie Haddon}, {and} \bibinfo{person}{Kjartan Olafsson}.} \bibinfo{year}{2009}\natexlab{}.
\newblock \bibinfo{booktitle}{\emph{Comparing children’s online opportunities and risks across Europe: Cross-national comparisons for EU Kids Online}}.
\newblock \bibinfo{publisher}{EU Kids Online, The London School of Economics and Political Science}.
\newblock
\showISBNx{ISBN 978-0-85328-406-2}
\urldef\tempurl%
\url{http://eprints.lse.ac.uk/24368/1/D3.2_Report-Cross_national_comparisons-2nd-edition.pdf}
\showURL{%
\tempurl}


\bibitem[Institute(2022)]%
        {Erikson_2022}
\bibfield{author}{\bibinfo{person}{Erikson Institute}.} \bibinfo{year}{2022}\natexlab{}.
\newblock \bibinfo{booktitle}{\emph{Technology and Young Children in the Digital Age}}.
\newblock \bibinfo{type}{{T}echnical {R}eport}. \bibinfo{institution}{Erikson Institute}.
\newblock
\urldef\tempurl%
\url{https://www.youthlead.org/resources/technology-and-young-children-digital-age}
\showURL{%
\tempurl}


\bibitem[Jackson et~al\mbox{.}(2014)]%
        {jackson2014}
\bibfield{author}{\bibinfo{person}{Steven~J. Jackson}, \bibinfo{person}{Tarleton Gillespie}, {and} \bibinfo{person}{Sandy Payette}.} \bibinfo{year}{2014}\natexlab{}.
\newblock \showarticletitle{The policy knot: re-integrating policy, practice and design in CSCW studies of social computing}. In \bibinfo{booktitle}{\emph{Proceedings of the 17th ACM Conference on Computer Supported Cooperative Work \& Social Computing}} (Baltimore, Maryland, USA) \emph{(\bibinfo{series}{CSCW '14})}. \bibinfo{publisher}{Association for Computing Machinery}, \bibinfo{address}{New York, NY, USA}, \bibinfo{pages}{588–602}.
\newblock
\showISBNx{9781450325400}
\urldef\tempurl%
\url{https://doi.org/10.1145/2531602.2531674}
\showDOI{\tempurl}


\bibitem[James et~al\mbox{.}(2019)]%
        {james2019teaching}
\bibfield{author}{\bibinfo{person}{Carrie James}, \bibinfo{person}{Emily Weinstein}, {and} \bibinfo{person}{Kelly Mendoza}.} \bibinfo{year}{2019}\natexlab{}.
\newblock \showarticletitle{Teaching digital citizens in today’s world: Research and insights behind the Common Sense K--12 Digital Citizenship Curriculum}.
\newblock \bibinfo{journal}{\emph{Common Sense Media}} (\bibinfo{year}{2019}), \bibinfo{pages}{2021--08}.
\newblock


\bibitem[Khan et~al\mbox{.}(2024)]%
        {10.1145/3613904.3642460}
\bibfield{author}{\bibinfo{person}{Sushmita Khan}, \bibinfo{person}{Mehtab Iqbal}, \bibinfo{person}{Oluwafemi Osho}, \bibinfo{person}{Khushbu Singh}, \bibinfo{person}{Kyra Derrick}, \bibinfo{person}{Philip Nelson}, \bibinfo{person}{Lingyuan Li}, \bibinfo{person}{Emily Sidnam-Mauch}, \bibinfo{person}{Nicole Bannister}, \bibinfo{person}{Kelly Caine}, {et~al\mbox{.}}} \bibinfo{year}{2024}\natexlab{}.
\newblock \showarticletitle{Teaching Middle Schoolers about the Privacy Threats of Tracking and Pervasive Personalization: A Classroom Intervention Using Design-Based Research}. In \bibinfo{booktitle}{\emph{Proceedings of the CHI Conference on Human Factors in Computing Systems}}. \bibinfo{pages}{1--26}.
\newblock
\urldef\tempurl%
\url{https://doi.org/10.1145/3613904.3642460}
\showDOI{\tempurl}


\bibitem[Kumar and Hyde(2023)]%
        {kumar2023exploring}
\bibfield{author}{\bibinfo{person}{Priya Kumar} {and} \bibinfo{person}{Lily Hyde}.} \bibinfo{year}{2023}\natexlab{}.
\newblock \showarticletitle{Exploring How US K-12 Education Addresses Privacy Literacy}.
\newblock \bibinfo{journal}{\emph{AoIR Selected Papers of Internet Research}} (\bibinfo{year}{2023}).
\newblock
\urldef\tempurl%
\url{https://doi.org/10.5210/spir.v2023i0.13439}
\showDOI{\tempurl}


\bibitem[Kumar et~al\mbox{.}(2017)]%
        {kumar2017no}
\bibfield{author}{\bibinfo{person}{Priya Kumar}, \bibinfo{person}{Shalmali~Milind Naik}, \bibinfo{person}{Utkarsha~Ramesh Devkar}, \bibinfo{person}{Marshini Chetty}, \bibinfo{person}{Tamara~L Clegg}, {and} \bibinfo{person}{Jessica Vitak}.} \bibinfo{year}{2017}\natexlab{}.
\newblock \showarticletitle{'No Telling Passcodes Out Because They're Private' Understanding Children's Mental Models of Privacy and Security Online}.
\newblock \bibinfo{journal}{\emph{Proceedings of the ACM on Human-Computer Interaction}} \bibinfo{volume}{1}, \bibinfo{number}{CSCW} (\bibinfo{year}{2017}), \bibinfo{pages}{1--21}.
\newblock
\urldef\tempurl%
\url{https://doi.org/10.1145/3134699}
\showDOI{\tempurl}


\bibitem[Kumar et~al\mbox{.}(2018)]%
        {kumar2018co}
\bibfield{author}{\bibinfo{person}{Priya Kumar}, \bibinfo{person}{Jessica Vitak}, \bibinfo{person}{Marshini Chetty}, \bibinfo{person}{Tamara~L. Clegg}, \bibinfo{person}{Jonathan Yang}, \bibinfo{person}{Brenna McNally}, {and} \bibinfo{person}{Elizabeth Bonsignore}.} \bibinfo{year}{2018}\natexlab{}.
\newblock \showarticletitle{Co-designing online privacy-related games and stories with children}. In \bibinfo{booktitle}{\emph{Proceedings of the 17th ACM Conference on Interaction Design and Children}} (Trondheim, Norway) \emph{(\bibinfo{series}{IDC '18})}. \bibinfo{publisher}{Association for Computing Machinery}, \bibinfo{address}{New York, NY, USA}, \bibinfo{pages}{67–79}.
\newblock
\showISBNx{9781450351522}
\urldef\tempurl%
\url{https://doi.org/10.1145/3202185.3202735}
\showDOI{\tempurl}


\bibitem[Kumar and Byrne(2022)]%
        {kumar20225ds}
\bibfield{author}{\bibinfo{person}{Priya~C Kumar} {and} \bibinfo{person}{Virginia~L Byrne}.} \bibinfo{year}{2022}\natexlab{}.
\newblock \showarticletitle{The 5Ds of privacy literacy: a framework for privacy education}.
\newblock \bibinfo{journal}{\emph{Information and Learning Sciences}} \bibinfo{volume}{123}, \bibinfo{number}{7/8} (\bibinfo{year}{2022}), \bibinfo{pages}{445--461}.
\newblock
\urldef\tempurl%
\url{https://doi.org/10.1108/ILS-02-2022-0022}
\showDOI{\tempurl}


\bibitem[Kumar et~al\mbox{.}(2019)]%
        {kumar2019privacy}
\bibfield{author}{\bibinfo{person}{Priya~C Kumar}, \bibinfo{person}{Marshini Chetty}, \bibinfo{person}{Tamara~L Clegg}, {and} \bibinfo{person}{Jessica Vitak}.} \bibinfo{year}{2019}\natexlab{}.
\newblock \showarticletitle{Privacy and security considerations for digital technology use in elementary schools}. In \bibinfo{booktitle}{\emph{Proceedings of the 2019 CHI Conference on Human Factors in Computing Systems}}. \bibinfo{pages}{1--13}.
\newblock
\urldef\tempurl%
\url{https://doi.org/10.1145/3290605.3300537}
\showDOI{\tempurl}


\bibitem[Kumar et~al\mbox{.}(2023)]%
        {kumar2023understanding}
\bibfield{author}{\bibinfo{person}{Priya~C Kumar}, \bibinfo{person}{Fiona O'Connell}, \bibinfo{person}{Lucy Li}, \bibinfo{person}{Virginia~L Byrne}, \bibinfo{person}{Marshini Chetty}, \bibinfo{person}{Tamara~L Clegg}, {and} \bibinfo{person}{Jessica Vitak}.} \bibinfo{year}{2023}\natexlab{}.
\newblock \showarticletitle{Understanding Research Related to Designing for Children's Privacy and Security: A Document Analysis}. In \bibinfo{booktitle}{\emph{Proceedings of the 22nd Annual ACM Interaction Design and Children Conference}}. \bibinfo{pages}{335--354}.
\newblock
\urldef\tempurl%
\url{https://doi.org/10.1145/3585088.3589375}
\showDOI{\tempurl}


\bibitem[Kumar et~al\mbox{.}(2020)]%
        {kumar2020strengthening}
\bibfield{author}{\bibinfo{person}{Priya~C Kumar}, \bibinfo{person}{Mega Subramaniam}, \bibinfo{person}{Jessica Vitak}, \bibinfo{person}{Tamara~L Clegg}, {and} \bibinfo{person}{Marshini Chetty}.} \bibinfo{year}{2020}\natexlab{}.
\newblock \showarticletitle{Strengthening children’s privacy literacy through contextual integrity}.
\newblock \bibinfo{journal}{\emph{Media and Communication}} \bibinfo{volume}{8}, \bibinfo{number}{4} (\bibinfo{year}{2020}), \bibinfo{pages}{175--184}.
\newblock
\urldef\tempurl%
\url{https://doi.org/10.17645/mac.v8i4.3236}
\showDOI{\tempurl}


\bibitem[Kyza and Nicolaidou(2017)]%
        {kyza2017co}
\bibfield{author}{\bibinfo{person}{Eleni~A Kyza} {and} \bibinfo{person}{Iolie Nicolaidou}.} \bibinfo{year}{2017}\natexlab{}.
\newblock \showarticletitle{Co-designing reform-based online inquiry learning environments as a situated approach to teachers’ professional development}.
\newblock \bibinfo{journal}{\emph{CoDesign}} \bibinfo{volume}{13}, \bibinfo{number}{4} (\bibinfo{year}{2017}), \bibinfo{pages}{261--286}.
\newblock
\urldef\tempurl%
\url{https://doi.org/10.1080/15710882.2016.1209528}
\showDOI{\tempurl}


\bibitem[Lamichhane and Read(2017)]%
        {lamichhane2017investigating}
\bibfield{author}{\bibinfo{person}{Dev~Raj Lamichhane} {and} \bibinfo{person}{Janet~C. Read}.} \bibinfo{year}{2017}\natexlab{}.
\newblock \showarticletitle{Investigating Children's Passwords using a Game-based Survey}. In \bibinfo{booktitle}{\emph{Proceedings of the 2017 Conference on Interaction Design and Children}} (Stanford, California, USA) \emph{(\bibinfo{series}{IDC '17})}. \bibinfo{publisher}{Association for Computing Machinery}, \bibinfo{address}{New York, NY, USA}, \bibinfo{pages}{617–622}.
\newblock
\showISBNx{9781450349215}
\urldef\tempurl%
\url{https://doi.org/10.1145/3078072.3084333}
\showDOI{\tempurl}


\bibitem[Lamond et~al\mbox{.}(2022)]%
        {lamond2022sok}
\bibfield{author}{\bibinfo{person}{Maria Lamond}, \bibinfo{person}{Karen Renaud}, \bibinfo{person}{Lara Wood}, {and} \bibinfo{person}{Suzanne Prior}.} \bibinfo{year}{2022}\natexlab{}.
\newblock \showarticletitle{SOK: young children's cybersecurity knowledge, skills \& practice: a systematic literature review}. In \bibinfo{booktitle}{\emph{Proceedings of the 2022 European Symposium on Usable Security}}. \bibinfo{pages}{14--27}.
\newblock
\urldef\tempurl%
\url{https://doi.org/10.1145/3549015.3554207}
\showDOI{\tempurl}


\bibitem[Lastdrager et~al\mbox{.}(2017)]%
        {lastdrager2017effective}
\bibfield{author}{\bibinfo{person}{Elmer Lastdrager}, \bibinfo{person}{In{\'e}s~Carvajal Gallardo}, \bibinfo{person}{Pieter Hartel}, {and} \bibinfo{person}{Marianne Junger}.} \bibinfo{year}{2017}\natexlab{}.
\newblock \showarticletitle{How Effective is $\{$Anti-Phishing$\}$ Training for Children?}. In \bibinfo{booktitle}{\emph{Thirteenth Symposium on Usable Privacy and Security (SOUPS 2017)}}. \bibinfo{publisher}{USENIX}, \bibinfo{pages}{229--239}.
\newblock
\urldef\tempurl%
\url{https://www.usenix.org/conference/soups2017/technical-sessions/presentation/lastdrager}
\showURL{%
\tempurl}


\bibitem[Li et~al\mbox{.}(2023)]%
        {li2023integrating}
\bibfield{author}{\bibinfo{person}{Yumeng Li}, \bibinfo{person}{Shaoshan Deng}, \bibinfo{person}{Xiaomin Wu}, \bibinfo{person}{Bin Zhao}, \bibinfo{person}{Yufei Xie}, \bibinfo{person}{Xianfei Luo}, {and} \bibinfo{person}{Yunxiang Zheng}.} \bibinfo{year}{2023}\natexlab{}.
\newblock \showarticletitle{Integrating Digital Citizenship into a Primary School Course “Ethics and the Rule of Law”: Necessity, Strategies and a Pilot Study}. In \bibinfo{booktitle}{\emph{International Conference on Blended Learning}}. Springer, \bibinfo{pages}{59--70}.
\newblock
\urldef\tempurl%
\url{https://doi.org/10.1007/978-3-031-35731-2_7}
\showDOI{\tempurl}


\bibitem[Liu et~al\mbox{.}(2024)]%
        {liu2024integrating}
\bibfield{author}{\bibinfo{person}{Lanjing Liu}, \bibinfo{person}{Lan Gao}, {and} \bibinfo{person}{Yaxing Yao}.} \bibinfo{year}{2024}\natexlab{}.
\newblock \showarticletitle{Integrating Family Privacy Education and Informal Learning Spaces: Characteristics, Challenges and Design Opportunities}. In \bibinfo{booktitle}{\emph{Extended Abstracts of the CHI Conference on Human Factors in Computing Systems}}. \bibinfo{pages}{1--9}.
\newblock
\urldef\tempurl%
\url{https://doi.org/10.1145/3613905.3650940}
\showDOI{\tempurl}


\bibitem[Livingstone(2006)]%
        {livingstone2006children}
\bibfield{author}{\bibinfo{person}{Sonia Livingstone}.} \bibinfo{year}{2006}\natexlab{}.
\newblock \bibinfo{booktitle}{\emph{Children’s Privacy Online: Experimenting with Boundaries Within and Beyond the Family}}.
\newblock \bibinfo{publisher}{Oxford University Press}, \bibinfo{pages}{128--144}.
\newblock
\showISBNx{978-0-19-531280-5}
\urldef\tempurl%
\url{https://doi.org/10.1093/acprof:oso/9780195312805.003.0010}
\showDOI{\tempurl}


\bibitem[Livingstone et~al\mbox{.}(2020)]%
        {livingstone2020data}
\bibfield{author}{\bibinfo{person}{Sonia Livingstone}, \bibinfo{person}{Mariya Stoilova}, {and} \bibinfo{person}{Rishita Nandagiri}.} \bibinfo{year}{2020}\natexlab{}.
\newblock \bibinfo{booktitle}{\emph{Data and Privacy Literacy}}.
\newblock \bibinfo{publisher}{John Wiley \& Sons, Ltd}, \bibinfo{pages}{413–425}.
\newblock
\showISBNx{978-1-119-16690-0}
\urldef\tempurl%
\url{https://doi.org/10.1002/9781119166900.ch38}
\showDOI{\tempurl}


\bibitem[Lu et~al\mbox{.}(2021)]%
        {lu2021coding}
\bibfield{author}{\bibinfo{person}{Alex~Jiahong Lu}, \bibinfo{person}{Gabriela Marcu}, \bibinfo{person}{Mark~S Ackerman}, {and} \bibinfo{person}{Tawanna~R Dillahunt}.} \bibinfo{year}{2021}\natexlab{}.
\newblock \showarticletitle{Coding bias in the use of behavior management technologies: Uncovering socio-technical consequences of data-driven surveillance in classrooms}. In \bibinfo{booktitle}{\emph{Proceedings of the 2021 ACM Designing Interactive Systems Conference}}. \bibinfo{pages}{508--522}.
\newblock
\urldef\tempurl%
\url{https://doi.org/10.1145/3461778.3462084}
\showDOI{\tempurl}


\bibitem[Maqsood et~al\mbox{.}(2018)]%
        {maqsood2018exploratory}
\bibfield{author}{\bibinfo{person}{Sumbal Maqsood}, \bibinfo{person}{Robert Biddle}, \bibinfo{person}{Sana Maqsood}, {and} \bibinfo{person}{Sonia Chiasson}.} \bibinfo{year}{2018}\natexlab{}.
\newblock \showarticletitle{An exploratory study of children's online password behaviours}. In \bibinfo{booktitle}{\emph{Proceedings of the 17th ACM Conference on Interaction Design and Children}} (Trondheim, Norway) \emph{(\bibinfo{series}{IDC '18})}. \bibinfo{publisher}{Association for Computing Machinery}, \bibinfo{address}{New York, NY, USA}, \bibinfo{pages}{539–544}.
\newblock
\showISBNx{9781450351522}
\urldef\tempurl%
\url{https://doi.org/10.1145/3202185.3210772}
\showDOI{\tempurl}


\bibitem[Maqsood and Chiasson(2021a)]%
        {maqsood2021design}
\bibfield{author}{\bibinfo{person}{Sana Maqsood} {and} \bibinfo{person}{Sonia Chiasson}.} \bibinfo{year}{2021}\natexlab{a}.
\newblock \showarticletitle{Design, development, and evaluation of a cybersecurity, privacy, and digital literacy game for tweens}.
\newblock \bibinfo{journal}{\emph{ACM Transactions on Privacy and Security (TOPS)}} \bibinfo{volume}{24}, \bibinfo{number}{4} (\bibinfo{year}{2021}), \bibinfo{pages}{1--37}.
\newblock
\urldef\tempurl%
\url{https://doi.org/10.1145/3469821}
\showDOI{\tempurl}


\bibitem[Maqsood and Chiasson(2021b)]%
        {maqsood2021they}
\bibfield{author}{\bibinfo{person}{Sana Maqsood} {and} \bibinfo{person}{Sonia Chiasson}.} \bibinfo{year}{2021}\natexlab{b}.
\newblock \showarticletitle{“They think it’s totally fine to talk to somebody on the internet they don’t know”: Teachers’ perceptions and mitigation strategies of tweens’ online risks}. In \bibinfo{booktitle}{\emph{Proceedings of the 2021 CHI Conference on Human Factors in Computing Systems}} (Yokohama, Japan) \emph{(\bibinfo{series}{CHI '21})}. \bibinfo{publisher}{Association for Computing Machinery}, \bibinfo{address}{New York, NY, USA}, Article \bibinfo{articleno}{688}, \bibinfo{numpages}{17}~pages.
\newblock
\showISBNx{9781450380966}
\urldef\tempurl%
\url{https://doi.org/10.1145/3411764.3445224}
\showDOI{\tempurl}


\bibitem[Martin et~al\mbox{.}(2023)]%
        {martin2023teacher}
\bibfield{author}{\bibinfo{person}{Florence Martin}, \bibinfo{person}{Julie Bacak}, \bibinfo{person}{Drew Polly}, \bibinfo{person}{Weichao Wang}, {and} \bibinfo{person}{Lynn Ahlgrim-Delzell}.} \bibinfo{year}{2023}\natexlab{}.
\newblock \showarticletitle{Teacher and School Concerns and Actions on Elementary School Children Digital Safety}.
\newblock \bibinfo{journal}{\emph{TechTrends}} \bibinfo{volume}{67}, \bibinfo{number}{3} (\bibinfo{year}{2023}), \bibinfo{pages}{561--571}.
\newblock
\urldef\tempurl%
\url{https://doi.org/10.1007/s11528-022-00803-z}
\showDOI{\tempurl}


\bibitem[McLeod(2023)]%
        {smith2023educators}
\bibfield{author}{\bibinfo{person}{Joy McLeod}.} \bibinfo{year}{2023}\natexlab{}.
\newblock \emph{\bibinfo{title}{Educators' Perspectives on Cybersecurity Educational Resources}}.
\newblock \bibinfo{thesistype}{Ph.\,D. Dissertation}. \bibinfo{school}{Carleton University}.
\newblock
\urldef\tempurl%
\url{https://doi.org/10.22215/etd/2023-15383}
\showDOI{\tempurl}


\bibitem[McLeod et~al\mbox{.}(2024)]%
        {mcleod2024comparing}
\bibfield{author}{\bibinfo{person}{Joy McLeod}, \bibinfo{person}{Leah Zhang-Kennedy}, {and} \bibinfo{person}{Elizabeth Stobert}.} \bibinfo{year}{2024}\natexlab{}.
\newblock \showarticletitle{Comparing Teacher and Creator Perspectives on the Design of Cybersecurity and Privacy Educational Resources}. In \bibinfo{booktitle}{\emph{Proceedings of the 2024 Symposium on Usable Privacy and Security (SOUPS 2024)}}. USENIX Association.
\newblock
\urldef\tempurl%
\url{https://www.usenix.org/conference/soups2024/presentation/mcleod}
\showURL{%
\tempurl}


\bibitem[McReynolds et~al\mbox{.}(2017)]%
        {mcreynolds2017toys}
\bibfield{author}{\bibinfo{person}{Emily McReynolds}, \bibinfo{person}{Sarah Hubbard}, \bibinfo{person}{Timothy Lau}, \bibinfo{person}{Aditya Saraf}, \bibinfo{person}{Maya Cakmak}, {and} \bibinfo{person}{Franziska Roesner}.} \bibinfo{year}{2017}\natexlab{}.
\newblock \showarticletitle{Toys that Listen: A Study of Parents, Children, and Internet-Connected Toys}. In \bibinfo{booktitle}{\emph{Proceedings of the 2017 CHI Conference on Human Factors in Computing Systems}} (Denver, Colorado, USA) \emph{(\bibinfo{series}{CHI '17})}. \bibinfo{publisher}{Association for Computing Machinery}, \bibinfo{address}{New York, NY, USA}, \bibinfo{pages}{5197–5207}.
\newblock
\showISBNx{9781450346559}
\urldef\tempurl%
\url{https://doi.org/10.1145/3025453.3025735}
\showDOI{\tempurl}


\bibitem[{Meta}(2024)]%
        {meta_youth_safety}
\bibfield{author}{\bibinfo{person}{{Meta}}.} \bibinfo{year}{2024}\natexlab{}.
\newblock \bibinfo{title}{Youth Safety}.
\newblock
\newblock
\urldef\tempurl%
\url{https://about.meta.com/actions/safety/audiences/youth}
\showURL{%
\tempurl}
\newblock
\shownote{Accessed: 2024-05}.


\bibitem[Nicholson et~al\mbox{.}(2020)]%
        {nicholson2020investigating}
\bibfield{author}{\bibinfo{person}{James Nicholson}, \bibinfo{person}{Yousra Javed}, \bibinfo{person}{Matt Dixon}, \bibinfo{person}{Lynne Coventry}, \bibinfo{person}{Opeyemi~Dele Ajayi}, {and} \bibinfo{person}{Philip Anderson}.} \bibinfo{year}{2020}\natexlab{}.
\newblock \showarticletitle{Investigating teenagers’ ability to detect phishing messages}. In \bibinfo{booktitle}{\emph{2020 IEEE European Symposium on Security and Privacy Workshops (EuroS\&PW)}}. IEEE, \bibinfo{pages}{140--149}.
\newblock
\urldef\tempurl%
\url{https://doi.org/10.1109/EuroSPW51379.2020.00027}
\showDOI{\tempurl}


\bibitem[Nicholson et~al\mbox{.}(2021)]%
        {nicholson2021understanding}
\bibfield{author}{\bibinfo{person}{James Nicholson}, \bibinfo{person}{Julia Terry}, \bibinfo{person}{Helen Beckett}, {and} \bibinfo{person}{Pardeep Kumar}.} \bibinfo{year}{2021}\natexlab{}.
\newblock \showarticletitle{Understanding young people's experiences of cybersecurity}. In \bibinfo{booktitle}{\emph{Proceedings of the 2021 European Symposium on Usable Security}}. \bibinfo{pages}{200--210}.
\newblock
\urldef\tempurl%
\url{https://doi.org/10.1145/3481357.3481520}
\showDOI{\tempurl}


\bibitem[Nissenbaum(2010)]%
        {Nissenbaum_2010}
\bibfield{author}{\bibinfo{person}{Helen Nissenbaum}.} \bibinfo{year}{2010}\natexlab{}.
\newblock \bibinfo{booktitle}{\emph{Privacy in Context: Technology, Policy, and the Integrity of Social Life}}.
\newblock \bibinfo{publisher}{Stanford University Press}.
\newblock
\showISBNx{978-0-8047-7289-1}


\bibitem[Nolan et~al\mbox{.}(2011)]%
        {nolan2011stranger}
\bibfield{author}{\bibinfo{person}{Jason Nolan}, \bibinfo{person}{Kate Raynes-Goldie}, {and} \bibinfo{person}{Melanie McBride}.} \bibinfo{year}{2011}\natexlab{}.
\newblock \showarticletitle{The Stranger Danger: Exploring Surveillance, Autonomy, and Privacy in Children's Use of Social Media.}
\newblock \bibinfo{journal}{\emph{Canadian Children}} \bibinfo{volume}{36}, \bibinfo{number}{2} (\bibinfo{year}{2011}).
\newblock
\urldef\tempurl%
\url{https://doi.org/10.18357/jcs.v36i2.15089}
\showDOI{\tempurl}


\bibitem[Norooz et~al\mbox{.}(2015)]%
        {norooz2015bodyvis}
\bibfield{author}{\bibinfo{person}{Leyla Norooz}, \bibinfo{person}{Matthew~Louis Mauriello}, \bibinfo{person}{Anita Jorgensen}, \bibinfo{person}{Brenna McNally}, {and} \bibinfo{person}{Jon~E Froehlich}.} \bibinfo{year}{2015}\natexlab{}.
\newblock \showarticletitle{BodyVis: A new approach to body learning through wearable sensing and visualization}. In \bibinfo{booktitle}{\emph{Proceedings of the 33rd Annual ACM Conference on Human Factors in Computing Systems}}. \bibinfo{pages}{1025--1034}.
\newblock
\urldef\tempurl%
\url{https://doi.org/10.1145/2702123.2702299}
\showDOI{\tempurl}


\bibitem[of~Education(2024)]%
        {ED_FERPA}
\bibfield{author}{\bibinfo{person}{U.S.~Department of Education}.} \bibinfo{year}{2024}\natexlab{}.
\newblock \bibinfo{title}{FERPA - Family Educational Rights and Privacy Act}.
\newblock
\newblock
\urldef\tempurl%
\url{https://www2.ed.gov/policy/gen/guid/fpco/ferpa/index.html}
\showURL{%
\tempurl}
\newblock
\shownote{Accessed: 2024-06}.


\bibitem[Palen and Dourish(2003)]%
        {palen2004}
\bibfield{author}{\bibinfo{person}{Leysia Palen} {and} \bibinfo{person}{Paul Dourish}.} \bibinfo{year}{2003}\natexlab{}.
\newblock \showarticletitle{Unpacking "privacy" for a networked world}. In \bibinfo{booktitle}{\emph{Proceedings of the SIGCHI Conference on Human Factors in Computing Systems}} (Ft. Lauderdale, Florida, USA) \emph{(\bibinfo{series}{CHI '03})}. \bibinfo{publisher}{Association for Computing Machinery}, \bibinfo{address}{New York, NY, USA}, \bibinfo{pages}{129–136}.
\newblock
\showISBNx{1581136307}
\urldef\tempurl%
\url{https://doi.org/10.1145/642611.642635}
\showDOI{\tempurl}


\bibitem[Pir et~al\mbox{.}(2023)]%
        {pir2023applying}
\bibfield{author}{\bibinfo{person}{Rumel MS~Rahman Pir}, \bibinfo{person}{Md~Forhad Rabbi}, {and} \bibinfo{person}{M~Jahirul Islam}.} \bibinfo{year}{2023}\natexlab{}.
\newblock \showarticletitle{Applying a machine learning model to forecast the risks to children's online privacy and security}. In \bibinfo{booktitle}{\emph{2023 International Conference on Intelligent Systems, Advanced Computing and Communication (ISACC)}}. IEEE, \bibinfo{pages}{1--8}.
\newblock
\urldef\tempurl%
\url{https://doi.org/10.1109/ISACC56298.2023.10084054}
\showDOI{\tempurl}


\bibitem[Quayyum(2020)]%
        {quayyum2020cyber}
\bibfield{author}{\bibinfo{person}{Farzana Quayyum}.} \bibinfo{year}{2020}\natexlab{}.
\newblock \showarticletitle{Cyber security education for children through gamification: research plan and perspectives}. In \bibinfo{booktitle}{\emph{Proceedings of the 2020 ACM Interaction Design and Children Conference: Extended Abstracts}}. \bibinfo{pages}{9--13}.
\newblock
\urldef\tempurl%
\url{https://doi.org/10.1145/3397617.3398030}
\showDOI{\tempurl}


\bibitem[Quayyum et~al\mbox{.}(2021)]%
        {quayyum2021cybersecurity}
\bibfield{author}{\bibinfo{person}{Farzana Quayyum}, \bibinfo{person}{Daniela~S Cruzes}, {and} \bibinfo{person}{Letizia Jaccheri}.} \bibinfo{year}{2021}\natexlab{}.
\newblock \showarticletitle{Cybersecurity awareness for children: A systematic literature review}.
\newblock \bibinfo{journal}{\emph{International Journal of Child-Computer Interaction}}  \bibinfo{volume}{30} (\bibinfo{year}{2021}), \bibinfo{pages}{100343}.
\newblock
\urldef\tempurl%
\url{https://doi.org/10.1016/j.ijcci.2021.100343}
\showDOI{\tempurl}


\bibitem[Rahman et~al\mbox{.}(2020)]%
        {rahman2020importance}
\bibfield{author}{\bibinfo{person}{Nurul Amirah~Abdul Rahman}, \bibinfo{person}{Izzah~Hanis Sairi}, \bibinfo{person}{Nurul Akma~M Zizi}, {and} \bibinfo{person}{Fariza Khalid}.} \bibinfo{year}{2020}\natexlab{}.
\newblock \showarticletitle{The importance of cybersecurity education in school}.
\newblock \bibinfo{journal}{\emph{International Journal of Information and Education Technology}} \bibinfo{volume}{10}, \bibinfo{number}{5} (\bibinfo{year}{2020}), \bibinfo{pages}{378--382}.
\newblock
\urldef\tempurl%
\url{https://doi.org/10.18178/ijiet.2020.10.5.1393}
\showDOI{\tempurl}


\bibitem[Raynes-Goldie and Allen(2014)]%
        {raynes2014gaming}
\bibfield{author}{\bibinfo{person}{Kate Raynes-Goldie} {and} \bibinfo{person}{Matthew Allen}.} \bibinfo{year}{2014}\natexlab{}.
\newblock \showarticletitle{Gaming privacy: A Canadian case study of a co-created privacy literacy game for children}.
\newblock  (\bibinfo{year}{2014}).
\newblock
\urldef\tempurl%
\url{https://hdl.handle.net/10536/DRO/DU:30065630}
\showURL{%
\tempurl}


\bibitem[Sa{\u{g}}lam et~al\mbox{.}(2023)]%
        {sauglam2023systematic}
\bibfield{author}{\bibinfo{person}{Rahime~Belen Sa{\u{g}}lam}, \bibinfo{person}{Vincent Miller}, {and} \bibinfo{person}{Virginia~NL Franqueira}.} \bibinfo{year}{2023}\natexlab{}.
\newblock \showarticletitle{A systematic literature review on cyber security education for children}.
\newblock \bibinfo{journal}{\emph{IEEE Transactions on Education}} \bibinfo{volume}{66}, \bibinfo{number}{3} (\bibinfo{year}{2023}), \bibinfo{pages}{274--286}.
\newblock
\urldef\tempurl%
\url{https://doi.org/10.1109/TE.2022.3231019}
\showDOI{\tempurl}


\bibitem[Salda{\~n}a(2021)]%
        {saldana2021coding}
\bibfield{author}{\bibinfo{person}{Johnny Salda{\~n}a}.} \bibinfo{year}{2021}\natexlab{}.
\newblock \bibinfo{booktitle}{\emph{The coding manual for qualitative researchers}}.
\newblock \bibinfo{publisher}{SAGE publications Ltd}.
\newblock


\bibitem[Sanders and Stappers(2008)]%
        {sanders2008co}
\bibfield{author}{\bibinfo{person}{Elizabeth B-N Sanders} {and} \bibinfo{person}{Pieter~Jan Stappers}.} \bibinfo{year}{2008}\natexlab{}.
\newblock \showarticletitle{Co-creation and the new landscapes of design}.
\newblock \bibinfo{journal}{\emph{Co-design}} \bibinfo{volume}{4}, \bibinfo{number}{1} (\bibinfo{year}{2008}), \bibinfo{pages}{5--18}.
\newblock
\urldef\tempurl%
\url{https://doi.org/10.1080/15710880701875068}
\showDOI{\tempurl}


\bibitem[Schwab et~al\mbox{.}(1992)]%
        {schwab1992}
\bibfield{author}{\bibinfo{person}{R.~G. Schwab}, \bibinfo{person}{Sylvia Hart-Landsberg}, \bibinfo{person}{Stephen Reder}, {and} \bibinfo{person}{Mark Abel}.} \bibinfo{year}{1992}\natexlab{}.
\newblock \showarticletitle{Collaboration and constraint: Middle school teaching teams}. In \bibinfo{booktitle}{\emph{Proceedings of the 1992 ACM Conference on Computer-Supported Cooperative Work}} (Toronto, Ontario, Canada) \emph{(\bibinfo{series}{CSCW '92})}. \bibinfo{publisher}{Association for Computing Machinery}, \bibinfo{address}{New York, NY, USA}, \bibinfo{pages}{241–248}.
\newblock
\showISBNx{0897915429}
\urldef\tempurl%
\url{https://doi.org/10.1145/143457.143513}
\showDOI{\tempurl}


\bibitem[Severance et~al\mbox{.}(2018)]%
        {severance2018organizing}
\bibfield{author}{\bibinfo{person}{Samuel Severance}, \bibinfo{person}{William~R Penuel}, \bibinfo{person}{Tamara Sumner}, {and} \bibinfo{person}{Heather Leary}.} \bibinfo{year}{2018}\natexlab{}.
\newblock \showarticletitle{Organizing for teacher agency in curricular co-design}.
\newblock In \bibinfo{booktitle}{\emph{Cultural-historical activity theory approaches to design-based research}}. \bibinfo{publisher}{Routledge}, \bibinfo{pages}{45--78}.
\newblock


\bibitem[Sevian et~al\mbox{.}(2018)]%
        {sevian2018does}
\bibfield{author}{\bibinfo{person}{Hannah Sevian}, \bibinfo{person}{Yehudit~Judy Dori}, {and} \bibinfo{person}{Ilka Parchmann}.} \bibinfo{year}{2018}\natexlab{}.
\newblock \showarticletitle{How does STEM context-based learning work: What we know and what we still do not know}.
\newblock \bibinfo{journal}{\emph{International Journal of Science Education}} \bibinfo{volume}{40}, \bibinfo{number}{10} (\bibinfo{year}{2018}), \bibinfo{pages}{1095--1107}.
\newblock
\urldef\tempurl%
\url{https://doi.org/10.1080/09500693.2018.1470346}
\showDOI{\tempurl}


\bibitem[Slov\'{a}k et~al\mbox{.}(2016)]%
        {Slovak2016}
\bibfield{author}{\bibinfo{person}{Petr Slov\'{a}k}, \bibinfo{person}{Kael Rowan}, \bibinfo{person}{Christopher Frauenberger}, \bibinfo{person}{Ran Gilad-Bachrach}, \bibinfo{person}{Mia Doces}, \bibinfo{person}{Brian Smith}, \bibinfo{person}{Rachel Kamb}, {and} \bibinfo{person}{Geraldine Fitzpatrick}.} \bibinfo{year}{2016}\natexlab{}.
\newblock \showarticletitle{Scaffolding the scaffolding: Supporting children's social-emotional learning at home}. In \bibinfo{booktitle}{\emph{Proceedings of the 19th ACM Conference on Computer-Supported Cooperative Work \& Social Computing}} (San Francisco, California, USA) \emph{(\bibinfo{series}{CSCW '16})}. \bibinfo{publisher}{Association for Computing Machinery}, \bibinfo{address}{New York, NY, USA}, \bibinfo{pages}{1751–1765}.
\newblock
\showISBNx{9781450335928}
\urldef\tempurl%
\url{https://doi.org/10.1145/2818048.2820007}
\showDOI{\tempurl}


\bibitem[Sobel et~al\mbox{.}(2017)]%
        {sobel2017}
\bibfield{author}{\bibinfo{person}{Kiley Sobel}, \bibinfo{person}{Geza Kovacs}, \bibinfo{person}{Galen McQuillen}, \bibinfo{person}{Andrew Cross}, \bibinfo{person}{Nirupama Chandrasekaran}, \bibinfo{person}{Nathalie~Henry Riche}, \bibinfo{person}{Ed Cutrell}, {and} \bibinfo{person}{Meredith~Ringel Morris}.} \bibinfo{year}{2017}\natexlab{}.
\newblock \showarticletitle{EduFeed: A Social Feed to Engage Preliterate Children in Educational Activities}. In \bibinfo{booktitle}{\emph{Proceedings of the 2017 ACM Conference on Computer Supported Cooperative Work and Social Computing}} (Portland, Oregon, USA) \emph{(\bibinfo{series}{CSCW '17})}. \bibinfo{publisher}{Association for Computing Machinery}, \bibinfo{address}{New York, NY, USA}, \bibinfo{pages}{491–504}.
\newblock
\showISBNx{9781450343350}
\urldef\tempurl%
\url{https://doi.org/10.1145/2998181.2998231}
\showDOI{\tempurl}


\bibitem[Team(2023)]%
        {nearpod2023digitalcitizenship}
\bibfield{author}{\bibinfo{person}{Nearpod Team}.} \bibinfo{year}{2023}\natexlab{}.
\newblock \bibinfo{title}{Digital Citizenship Week: Free Lessons and Activities for K-12}.
\newblock
\newblock
\urldef\tempurl%
\url{https://nearpod.com/blog/digital-citizenship-week-free-lessons/}
\showURL{%
\tempurl}
\newblock
\shownote{Accessed: 2024-06}.


\bibitem[Tirumala et~al\mbox{.}(2016)]%
        {tirumala2016survey}
\bibfield{author}{\bibinfo{person}{Sreenivas~Sremath Tirumala}, \bibinfo{person}{Abdolhossein Sarrafzadeh}, {and} \bibinfo{person}{Paul Pang}.} \bibinfo{year}{2016}\natexlab{}.
\newblock \showarticletitle{A survey on internet usage and cybersecurity awareness in students}. In \bibinfo{booktitle}{\emph{2016 14th Annual Conference on Privacy, Security and Trust (PST)}}. IEEE, \bibinfo{pages}{223--228}.
\newblock
\urldef\tempurl%
\url{https://doi.org/10.1109/PST.2016.7906931}
\showDOI{\tempurl}


\bibitem[\v{S}v\'{a}bensk\'{y} et~al\mbox{.}(2020)]%
        {vsvabensky2020cybersecurity}
\bibfield{author}{\bibinfo{person}{Valdemar \v{S}v\'{a}bensk\'{y}}, \bibinfo{person}{Jan Vykopal}, {and} \bibinfo{person}{Pavel \v{C}eleda}.} \bibinfo{year}{2020}\natexlab{}.
\newblock \showarticletitle{What Are Cybersecurity Education Papers About? A Systematic Literature Review of SIGCSE and ITiCSE Conferences} \emph{(\bibinfo{series}{SIGCSE '20})}. \bibinfo{publisher}{Association for Computing Machinery}, \bibinfo{address}{New York, NY, USA}, \bibinfo{pages}{2–8}.
\newblock
\showISBNx{9781450367936}
\urldef\tempurl%
\url{https://doi.org/10.1145/3328778.3366816}
\showDOI{\tempurl}


\bibitem[Wagman et~al\mbox{.}(2023)]%
        {wagman2023we}
\bibfield{author}{\bibinfo{person}{Kelly~B Wagman}, \bibinfo{person}{Elana~B Blinder}, \bibinfo{person}{Kevin Song}, \bibinfo{person}{Antoine Vignon}, \bibinfo{person}{Solomon Dworkin}, \bibinfo{person}{Tamara Clegg}, \bibinfo{person}{Jessica Vitak}, {and} \bibinfo{person}{Marshini Chetty}.} \bibinfo{year}{2023}\natexlab{}.
\newblock \showarticletitle{``We picked community over privacy''': Privacy and Security Concerns Emerging from Remote Learning Sociotechnical Infrastructure During COVID-19}.
\newblock \bibinfo{journal}{\emph{Proceedings of the ACM on Human-Computer Interaction}} \bibinfo{volume}{7}, \bibinfo{number}{CSCW2} (\bibinfo{year}{2023}), \bibinfo{pages}{1--29}.
\newblock
\urldef\tempurl%
\url{https://doi.org/10.1145/3610036}
\showDOI{\tempurl}


\bibitem[Wang et~al\mbox{.}(2024)]%
        {ge2024koala}
\bibfield{author}{\bibinfo{person}{Ge Wang}, \bibinfo{person}{Jun Zhao}, \bibinfo{person}{Konrad Kollnig}, \bibinfo{person}{Adrien Zier}, \bibinfo{person}{Blanche Duron}, \bibinfo{person}{Zhilin Zhang}, \bibinfo{person}{Max Van~Kleek}, {and} \bibinfo{person}{Nigel Shadbolt}.} \bibinfo{year}{2024}\natexlab{}.
\newblock \showarticletitle{KOALA Hero Toolkit: A New Approach to Inform Families of Mobile Datafication Risks}. In \bibinfo{booktitle}{\emph{Proceedings of the CHI Conference on Human Factors in Computing Systems}} (Honolulu, HI, USA) \emph{(\bibinfo{series}{CHI '24})}. \bibinfo{publisher}{Association for Computing Machinery}, \bibinfo{address}{New York, NY, USA}, Article \bibinfo{articleno}{226}, \bibinfo{numpages}{18}~pages.
\newblock
\showISBNx{9798400703300}
\urldef\tempurl%
\url{https://doi.org/10.1145/3613904.3642283}
\showDOI{\tempurl}


\bibitem[Wang et~al\mbox{.}(2022)]%
        {ge2022dont}
\bibfield{author}{\bibinfo{person}{Ge Wang}, \bibinfo{person}{Jun Zhao}, \bibinfo{person}{Max Van~Kleek}, {and} \bibinfo{person}{Nigel Shadbolt}.} \bibinfo{year}{2022}\natexlab{}.
\newblock \showarticletitle{'Don't make assumptions about me!': Understanding Children's Perception of Datafication Online}.
\newblock \bibinfo{journal}{\emph{Proc. ACM Hum.-Comput. Interact.}} \bibinfo{volume}{6}, \bibinfo{number}{CSCW2}, Article \bibinfo{articleno}{419} (\bibinfo{date}{nov} \bibinfo{year}{2022}), \bibinfo{numpages}{24}~pages.
\newblock
\urldef\tempurl%
\url{https://doi.org/10.1145/3555144}
\showDOI{\tempurl}


\bibitem[Wang et~al\mbox{.}(2023)]%
        {ge2022treat}
\bibfield{author}{\bibinfo{person}{Ge Wang}, \bibinfo{person}{Jun Zhao}, \bibinfo{person}{Max Van~Kleek}, {and} \bibinfo{person}{Nigel Shadbolt}.} \bibinfo{year}{2023}\natexlab{}.
\newblock \showarticletitle{‘Treat me as your friend, not a number in your database’: Co-designing with Children to Cope with Datafication Online}. In \bibinfo{booktitle}{\emph{Proceedings of the 2023 CHI Conference on Human Factors in Computing Systems}} (Hamburg, Germany) \emph{(\bibinfo{series}{CHI '23})}. \bibinfo{publisher}{Association for Computing Machinery}, \bibinfo{address}{New York, NY, USA}, Article \bibinfo{articleno}{95}, \bibinfo{numpages}{21}~pages.
\newblock
\showISBNx{9781450394215}
\urldef\tempurl%
\url{https://doi.org/10.1145/3544548.3580933}
\showDOI{\tempurl}


\bibitem[Williams et~al\mbox{.}(2023)]%
        {williams2023youth}
\bibfield{author}{\bibinfo{person}{Olivia Williams}, \bibinfo{person}{Yee-Yin Choong}, {and} \bibinfo{person}{Kerrianne Buchanan}.} \bibinfo{year}{2023}\natexlab{}.
\newblock \showarticletitle{Youth understandings of online privacy and security: A dyadic study of children and their parents}. In \bibinfo{booktitle}{\emph{Nineteenth Symposium on Usable Privacy and Security (SOUPS 2023)}}. \bibinfo{pages}{399--416}.
\newblock
\urldef\tempurl%
\url{https://www.usenix.org/conference/soups2023/presentation/williams}
\showURL{%
\tempurl}


\bibitem[Wisniewski et~al\mbox{.}(2015)]%
        {wisniewski2015}
\bibfield{author}{\bibinfo{person}{Pamela Wisniewski}, \bibinfo{person}{Haiyan Jia}, \bibinfo{person}{Heng Xu}, \bibinfo{person}{Mary~Beth Rosson}, {and} \bibinfo{person}{John~M. Carroll}.} \bibinfo{year}{2015}\natexlab{}.
\newblock \showarticletitle{"Preventative" vs. "Reactive": How Parental Mediation Influences Teens' Social Media Privacy Behaviors}. In \bibinfo{booktitle}{\emph{Proceedings of the 18th ACM Conference on Computer Supported Cooperative Work \& Social Computing}} (Vancouver, BC, Canada) \emph{(\bibinfo{series}{CSCW '15})}. \bibinfo{publisher}{Association for Computing Machinery}, \bibinfo{address}{New York, NY, USA}, \bibinfo{pages}{302–316}.
\newblock
\showISBNx{9781450329224}
\urldef\tempurl%
\url{https://doi.org/10.1145/2675133.2675293}
\showDOI{\tempurl}


\bibitem[Yan et~al\mbox{.}(2021)]%
        {yan2021risk}
\bibfield{author}{\bibinfo{person}{Zheng Yan}, \bibinfo{person}{Yukang Xue}, {and} \bibinfo{person}{Yaosheng Lou}.} \bibinfo{year}{2021}\natexlab{}.
\newblock \showarticletitle{Risk and protective factors for intuitive and rational judgment of cybersecurity risks in a large sample of K-12 students and teachers}.
\newblock \bibinfo{journal}{\emph{Computers in Human Behavior}}  \bibinfo{volume}{121} (\bibinfo{year}{2021}), \bibinfo{pages}{106791}.
\newblock
\urldef\tempurl%
\url{https://doi.org/10.1016/j.chb.2021.106791}
\showDOI{\tempurl}


\bibitem[Yap and Lee(2020)]%
        {yap2020phone}
\bibfield{author}{\bibinfo{person}{Christine Ee~Ling Yap} {and} \bibinfo{person}{Jung-Joo Lee}.} \bibinfo{year}{2020}\natexlab{}.
\newblock \showarticletitle{'Phone apps know a lot about you!': educating early adolescents about informational privacy through a phygital interactive book}. In \bibinfo{booktitle}{\emph{Proceedings of the Interaction Design and Children Conference}} (London, United Kingdom) \emph{(\bibinfo{series}{IDC '20})}. \bibinfo{publisher}{Association for Computing Machinery}, \bibinfo{address}{New York, NY, USA}, \bibinfo{pages}{49–62}.
\newblock
\showISBNx{9781450379816}
\urldef\tempurl%
\url{https://doi.org/10.1145/3392063.3394420}
\showDOI{\tempurl}


\bibitem[Yu et~al\mbox{.}(2015)]%
        {yu2015enhancing}
\bibfield{author}{\bibinfo{person}{Kuang-Chao Yu}, \bibinfo{person}{Szu-Chun Fan}, {and} \bibinfo{person}{Kuen-Yi Lin}.} \bibinfo{year}{2015}\natexlab{}.
\newblock \showarticletitle{Enhancing Students’ Problem-Solving Skills Through Context-Based Learning}.
\newblock \bibinfo{journal}{\emph{International Journal of Science and Mathematics Education}} \bibinfo{volume}{13}, \bibinfo{number}{6} (\bibinfo{date}{Dec.} \bibinfo{year}{2015}), \bibinfo{pages}{1377–1401}.
\newblock
\showISSN{1573-1774}
\urldef\tempurl%
\url{https://doi.org/10.1007/s10763-014-9567-4}
\showDOI{\tempurl}


\bibitem[Zhang-Kennedy et~al\mbox{.}(2017)]%
        {zhang2017cyberheroes}
\bibfield{author}{\bibinfo{person}{Leah Zhang-Kennedy}, \bibinfo{person}{Yomna Abdelaziz}, {and} \bibinfo{person}{Sonia Chiasson}.} \bibinfo{year}{2017}\natexlab{}.
\newblock \showarticletitle{Cyberheroes: The design and evaluation of an interactive ebook to educate children about online privacy}.
\newblock \bibinfo{journal}{\emph{International Journal of Child-Computer Interaction}}  \bibinfo{volume}{13} (\bibinfo{year}{2017}), \bibinfo{pages}{10--18}.
\newblock
\urldef\tempurl%
\url{https://doi.org/10.1016/j.ijcci.2017.05.001}
\showDOI{\tempurl}


\bibitem[Zhang-Kennedy and Chiasson(2021)]%
        {zhang2021systematic}
\bibfield{author}{\bibinfo{person}{Leah Zhang-Kennedy} {and} \bibinfo{person}{Sonia Chiasson}.} \bibinfo{year}{2021}\natexlab{}.
\newblock \showarticletitle{A systematic review of multimedia tools for cybersecurity awareness and education}.
\newblock \bibinfo{journal}{\emph{ACM Computing Surveys (CSUR)}} \bibinfo{volume}{54}, \bibinfo{number}{1} (\bibinfo{year}{2021}), \bibinfo{pages}{1--39}.
\newblock
\urldef\tempurl%
\url{https://doi.org/10.1145/3427920}
\showDOI{\tempurl}


\bibitem[Zhang-Kennedy et~al\mbox{.}(2016)]%
        {zhang2016nosy}
\bibfield{author}{\bibinfo{person}{Leah Zhang-Kennedy}, \bibinfo{person}{Christine Mekhail}, \bibinfo{person}{Yomna Abdelaziz}, {and} \bibinfo{person}{Sonia Chiasson}.} \bibinfo{year}{2016}\natexlab{}.
\newblock \showarticletitle{From Nosy Little Brothers to Stranger-Danger: Children and Parents' Perception of Mobile Threats}. In \bibinfo{booktitle}{\emph{Proceedings of the The 15th International Conference on Interaction Design and Children}} (Manchester, United Kingdom) \emph{(\bibinfo{series}{IDC '16})}. \bibinfo{publisher}{Association for Computing Machinery}, \bibinfo{address}{New York, NY, USA}, \bibinfo{pages}{388–399}.
\newblock
\showISBNx{9781450343138}
\urldef\tempurl%
\url{https://doi.org/10.1145/2930674.2930716}
\showDOI{\tempurl}


\bibitem[Zhao et~al\mbox{.}(2022)]%
        {zhao2022koala}
\bibfield{author}{\bibinfo{person}{Jun Zhao}, \bibinfo{person}{Blanche Duron}, {and} \bibinfo{person}{Ge Wang}.} \bibinfo{year}{2022}\natexlab{}.
\newblock \showarticletitle{KOALA Hero: Inform Children of Privacy Risks of Mobile Apps}. In \bibinfo{booktitle}{\emph{Proceedings of the 21st Annual ACM Interaction Design and Children Conference}} (Braga, Portugal) \emph{(\bibinfo{series}{IDC '22})}. \bibinfo{publisher}{Association for Computing Machinery}, \bibinfo{address}{New York, NY, USA}, \bibinfo{pages}{523–528}.
\newblock
\showISBNx{9781450391979}
\urldef\tempurl%
\url{https://doi.org/10.1145/3501712.3535278}
\showDOI{\tempurl}


\bibitem[Zhao et~al\mbox{.}(2019)]%
        {zhao2019make}
\bibfield{author}{\bibinfo{person}{Jun Zhao}, \bibinfo{person}{Ge Wang}, \bibinfo{person}{Carys Dally}, \bibinfo{person}{Petr Slovak}, \bibinfo{person}{Julian Edbrooke-Childs}, \bibinfo{person}{Max Van~Kleek}, {and} \bibinfo{person}{Nigel Shadbolt}.} \bibinfo{year}{2019}\natexlab{}.
\newblock \showarticletitle{`I make up a silly name': Understanding Children's Perception of Privacy Risks Online}. In \bibinfo{booktitle}{\emph{Proceedings of the 2019 CHI Conference on Human Factors in Computing Systems}} (Glasgow, Scotland Uk) \emph{(\bibinfo{series}{CHI '19})}. \bibinfo{publisher}{Association for Computing Machinery}, \bibinfo{address}{New York, NY, USA}, \bibinfo{pages}{1–13}.
\newblock
\showISBNx{9781450359702}
\urldef\tempurl%
\url{https://doi.org/10.1145/3290605.3300336}
\showDOI{\tempurl}


\end{thebibliography}

\appendix
\newpage
\section{Appendix}

\subsection{Formative Study Protocol}
\label{sec:formativeprotocol}

\subsubsection{Opening}
Welcome and thanks for joining us for this study. [Moderator self-introduction]. Today we’ll be talking about how students use technology in the classroom and in the home, and any challenges they have encountered you are aware of. We’ll also discuss your perceptions of children’s attitudes toward privacy and security, as well as your experiences helping children navigate privacy and security online. 

\textit{(Group Interview Only)} The format of this session is a focus group. I have a set of questions I’d like to open up to discussion, but there’s no formal method for answering. I encourage everyone to share their thoughts. My role is merely to facilitate the conversation; you all will be guiding it. 

\textit{(Individual Interview Only)} The format of this session is a 1-1 interview. I have a set of questions I’d like to open up to ask, but there’s no formal method for answering. I encourage you to share your thoughts. 

This session is scheduled to last approximately 60 minutes. Does anyone have questions before we start? Can I also record our conversation for today? I want to assure you that whatever that is being shared in this room today stays with us, and anything we use from this conversation today to develop resources or publications will be using pseudonyms. 

\subsubsection{Warm-up Activity}
Let’s start with a quick warm-up activity. 
\begin{itemize}
    \item Could you share your name, what subject, and what grade you teach? 
\end{itemize}

\subsubsection{Children’s Technology Use, Privacy, and Security in Remote Learning}
Let’s get started by talking a bit about the technology you use in the (remote) classroom. 
\begin{itemize}
    \item Do you think kids have a good understanding of what privacy means, both on and off the Internet? 
    \item Do you know how technology and social media are used in the home? 
    \item Do you think parents are usually aware of what kids are doing on their devices and on the Internet? 
\end{itemize}

\subsubsection{Teaching Digital Privacy and Security in the Classroom}
\begin{itemize}
    \item Do you think students are taught the importance of computer privacy and security at home? 
    \item What are some issues you have experienced with kids not understanding privacy? For example, a case where a student did not keep another student’s information private? 
    \item What are some ways you’ve tried to resolve these issues? How well did they work? 
    \item What are some methods you have tried in the past to teach kids about computer security and data privacy, if any at all? 
\end{itemize}

\subsubsection{Designing Digital Privacy and Security Lessons}
\begin{itemize}
    \item What do you hope kids learn about computer security and data privacy from the lesson plans we develop together? 
    \item What is, in your opinion, the most important thing that students should understand about privacy and be involved in the lesson plans we develop? 
\end{itemize}

\subsubsection{Ending}
Thank you again for your time today, which we know is very valuable. We appreciate your contributions and will be happy to share results from this project with anyone who is interested.
\clearpage

\newpage
\subsection{Evaluation Interview Protocol\protect\footnote{We only present the protocol used for the lesson design initial evaluation. The protocol used for the final evaluation has the same structure but tweaked question descriptions. }}
\label{sec:evaluationprotocol}
\subsubsection{Begining}
 
Thanks for taking the time to join us to talk to us about your experiences implementing the micro-lesson series. [Interviewers self-introduction]. 
 
This interview will last for approximately 60 minutes. During this time we’ll be asking you to respond to questions about your experience teaching with the curricular privacy and security resources and your perspectives about how they might be improved. We’re really appreciative of your willingness to speak with us, but I do want to remind you that you do not need to respond to any questions you’d prefer to skip and that you are free to withdraw from the study at any time.
 
We’ll be video recording and transcribing this session. This recording is for note-taking purposes only and will not be shared with anybody outside of our research team. We won't use your real name or any other potentially identifying information in any related publications.
 
Before I begin recording, do you have any questions about the consent information or about the interview, or anything else I’ve just shared? If you consent to us recording the interview, I’ll start that recording now. 

\subsubsection{Background Information}
To begin, I’d like to ask you a few general questions to get a sense of you and your teaching experience.

\begin{itemize}
    \item Could you please begin by confirming your name and the grade(s) and subject(s) you currently teach?
    \item And how many years have you been teaching in this grade/subject area and overall?
    \item What motivated you to participate in this study?
    \item Have you ever taught your students about online privacy and security related topics before implementing the micro-lesson series?
    \begin{itemize}
        \item Tell me about it. What did you do and how did it go?
    \end{itemize}
\end{itemize}

\subsubsection{Experiences Implementing Micro-Lessons}
Now we’d like to move on to discuss your experiences using the privacy and security resources and activities we provided with your students.
\begin{itemize}
    \item Tell us about your overall experience implementing these lessons. What did you do?
    \begin{itemize}
        \item Please share any adaptations and/or additions you made to the provided micro-lessons and why.
        \item Do you have documented plans you can share/walk us through?
    \end{itemize}
    \item Tell me your thoughts about teaching the privacy and security concepts and content covered in the lessons.
    \begin{itemize}
        \item How comfortable/confident were you with the content?
        \item How knowledgeable were you about the content?
        \item Were there any specific topics that you felt more or less comfortable teaching? Why?
    \end{itemize}
\end{itemize}
Now I want to ask you some questions about what was helpful and what was challenging about using these resources in your classroom, for you and for your students.
\begin{itemize}
    \item For you, as a teacher, what was:
    \begin{itemize}
        \item most helpful about using these resources?
        \item most challenging about using these resources?
    \end{itemize}

    \item For your students participating in these lessons, what was:
    \begin{itemize}
        \item helpful about these lessons and activities?
        \item challenging about these lessons and activities?
        \item (If not already addressed) How relevant was the content for your students?
        \item (If not already addressed) How relevant and accessible were the activities to your students?
    \end{itemize}
    
    \item How much do you think your students were engaged in these lessons overall?
    \begin{itemize}
        \item Were there any memorable moments that stand out?
        \item Were there any lessons where students appeared especially engaged or disengaged?
        \item Were there any groups of students that were particularly engaged or disengaged? Tell us about them.
    \end{itemize}
\end{itemize}

\subsubsection{Takeaways from Micro-Lessons}
\begin{itemize}
    \item How much do you think your students learned from these lessons overall?
    \begin{itemize}
        \item What do you feel your students’ main takeaways were from the lesson series?
        \item To what extent did you feel the lessons were accessible to any students with special needs in your classroom?
        \item Are there any examples of student work that stood out to you from the lesson series?
    \end{itemize}

    \item (If not already addressed) Were there any lessons, materials, or activities that stood out to you as particularly effective? What was it about these lessons that made them successful?

    \item (If not already addressed) Were there any lessons, materials, or activities that stood out to you as particularly ineffective for your students? What was it about these lessons that made them less successful?

    \item What did you, as a teacher, learn from implementing these lessons with your students?
    \begin{itemize}
        \item What materials and/or activities supported your learning about privacy and security topics?
    \end{itemize}
\end{itemize}

\subsubsection{Recommendations of Improving Micro-Lessons}
\begin{itemize}
    \item What recommendations do you have to improve these lessons and resources?
    \begin{itemize}
        \item broadly?
        \item for specific groups of students (e.g., special needs)?
        \item Are there any topics you think should have been covered that weren’t?
    \end{itemize}

    \item What recommendations do you have to improve teacher learning about these topics?
\end{itemize}

\subsubsection{Ending}
\begin{itemize}
    \item Finally, I’d like to ask you if there’s anything I haven’t asked you, that you think is important to share? Or are there any other thoughts you’d like to share with us before we conclude the interview? 

    \item I’d also like to ask the [second researcher] if there are any follow-up questions you’d like to ask before we conclude the interview.
\end{itemize}

Thank you again for your time today, which we know is very valuable. We appreciate your contributions and will be happy to share results from this project with anyone who is interested.

\clearpage

\newpage
\subsection{Codebook for Data Analysis}
\label{sec:appcode}
\begin{table}[h]
\centering
\renewcommand\arraystretch{1.5}
\caption{Codebook for formative study data analysis with description for each sub-code. Structural codes are bolded. }
\label{tab:formativecode}
\resizebox{\linewidth}{!}{
\begin{tabular}{p{0.35\textwidth}p{0.65\textwidth}} 
\toprule
\textbf{Code}                                              & \textbf{Description}                                                                                                               \\ 
\midrule
\textbf{Digital Privacy and Security Concepts for Children}             &                                                                                                                                    \\
Privacy and security concerns                              & Reports of concerns of children's privacy and security                                                                             \\
Privacy and security incidents                             & Reports of privacy or security related incidents with students at school or at home                                                \\
Digital literacy meanings                                   & Teacher's definition of digital literacy for children                                                                              \\ 
\midrule
\textbf{Privacy and Security Curriculum}                   &                                                                                                                                    \\
Current privacy and security teaching                      & Reports of teachers taught/didn't taught privacy and security in their class, and comments on current approaches                   \\
Curriculum -- barriers                                      & Reports of barriers in incorporating privacy and security teaching/lessons in their class                                          \\ 
\midrule
\textbf{Professional Development (PD)}                          &                                                                                                                                    \\
PD -- suggestions                                           & Reports of suggestions to provide useful privacy and security training for teachers                                       \\
Current PD                           & Reports of teachers took/didn't take professional development related to privacy and security, and comments on current approaches  \\
PD -- challenges                                            & Reports of challenges of implementing personal development.                                                                        \\
\midrule
\textbf{Designing Digital Privacy and Security Micro-Lessons} &                                                                                                                                    \\
Important content                                          & Report of important skills and knowledge for children to learn about privacy and security                                  \\
Micro-Lessons -- suggestions                                & Reports of suggestions for children to learn about privacy and security at school                                                  \\
Micro-Lessons                                              & Reports of benefits of mini-lessons rather than one big lesson to teach children privacy and security                              \\ 

\bottomrule
\end{tabular}}
\end{table}

\begin{table}[h]
\centering
\renewcommand\arraystretch{1.5}
\caption{Codebook for evaluation sessions data analysis with description for each sub-code. Structural codes are bolded.}
\label{tab:evaluationcode}
\resizebox{\linewidth}{!}{
\begin{tabular}{p{0.35\textwidth}p{0.65\textwidth}} 
\toprule
\textbf{Code}                                           & \textbf{Description}                                                                                                        \\ 
\midrule
\textbf{Class Implementation Approaches}                 &                                                                                                                             \\
General information                                      & Reports of the overall implementation of the micro-lessons                                                                  \\
Addition and adaptation in teaching -- aim                & Reports of reasons for teachers adding to/modifying the micro-lesson content in class implementation                        \\
Addition and adaptation in teaching -- process            & Reports of actions for teachers adding to/modifying the micro-lesson content                                                \\
Addition and adaptation in teaching -- resources          & Reports of resources/references for teachers adding to/modifying micro-lesson content                                       \\ 
\midrule
\textbf{Strengths and Challenges Teacher Teaching}       &                                                                                                                             \\
Teacher’s satisfaction of micro-lesson                   & Reports of lesson plan components that teachers were satisfied with and facilitate class preparation and implementation     \\
Teacher's dissatisfaction of micro-lesson                & Reports of lesson plan components that teachers were dissatisifed with and need to be improved for better teaching process  \\
Teacher's takeaways from teaching                        & Reports of things teachers learned from preparing and implementing micro-lessons                                            \\
Challenges during teaching                               & Reports of challenges for teachers preparing and implementing micro-lessons                                                 \\ 
\midrule
\textbf{Opportunities and Difficulties Student Learning} &                                                                                                                             \\
Advantage of lesson for student learning                 & Discussions of lesson content and activities that were beneficial for students                                              \\
Student learning challenges and risks                    & Discussions of difficulties for students to engage in and learn from micro-lessons                                          \\
Student learning outcome                                 & Discussions of things students learned / progress students made in privacy and security after taking micro-lessons  \\ 
\hline
\textbf{Potential Lesson Improvement}                    &                                                                                                                             \\
Course materials and activities for students             & Discussions of methods to improve lesson contents and activities for students                                               \\
Assistance and resources for teachers                    & Discussions of methods to improving teacher resources                                                                       \\
Teaching strategy recommendation                         & Discussions of methods for adjusting teaching arrangements and strategies                                                    \\
\bottomrule
\end{tabular}}
\end{table}
\clearpage

\newpage
\subsection{Sample of Final Lesson Plan Design}
\label{sec:finaldesign}

\begin{figure}[h]
    \centering
    \includegraphics[width= 0.7\textwidth]{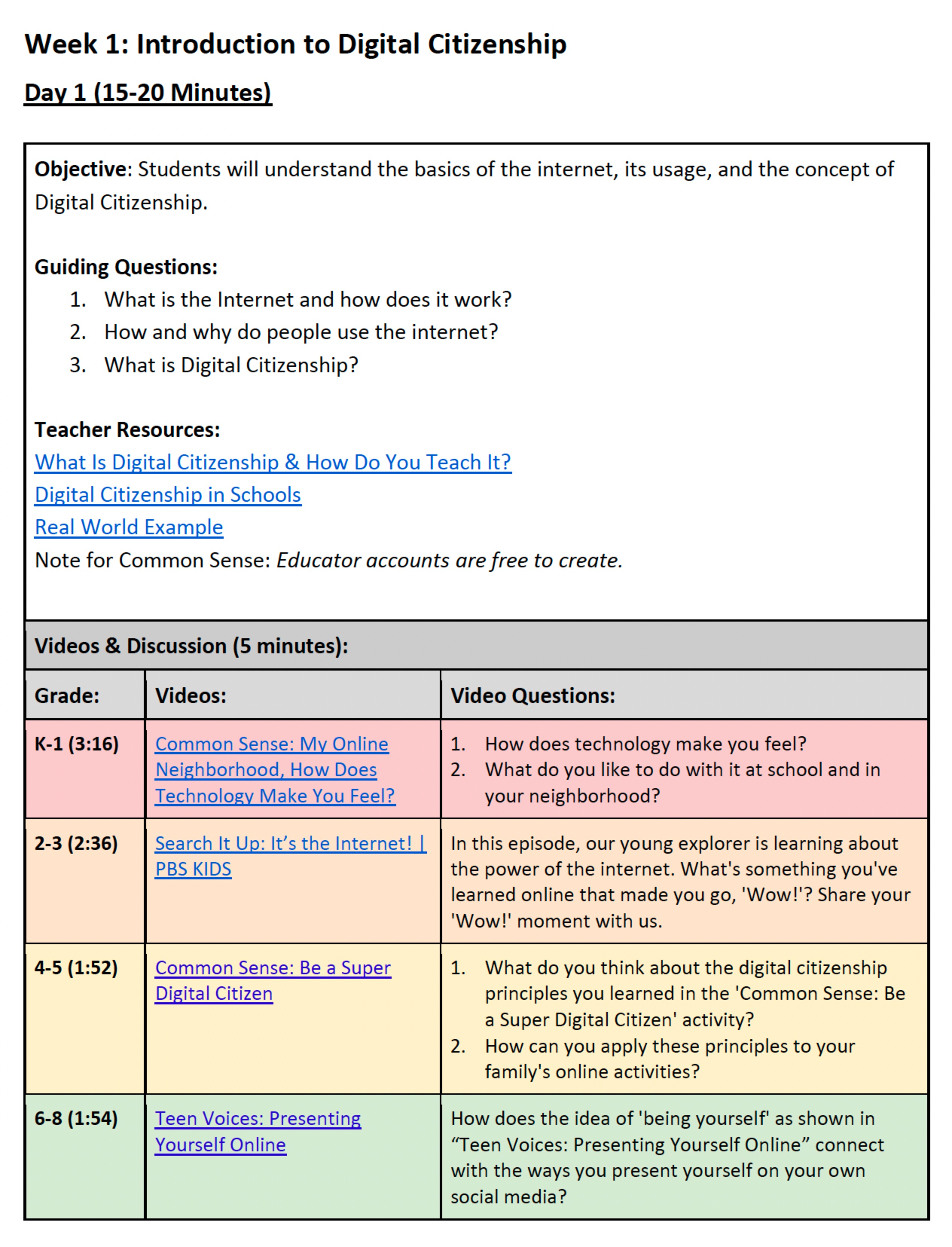}
    \caption{Final lesson plan design for lesson 1 (day 1) in module 1: Digital Citizenship.}
    \label{fig:finallesson1}
\end{figure}

\begin{figure}[h]
    \centering
    \includegraphics[width= 0.7\textwidth]{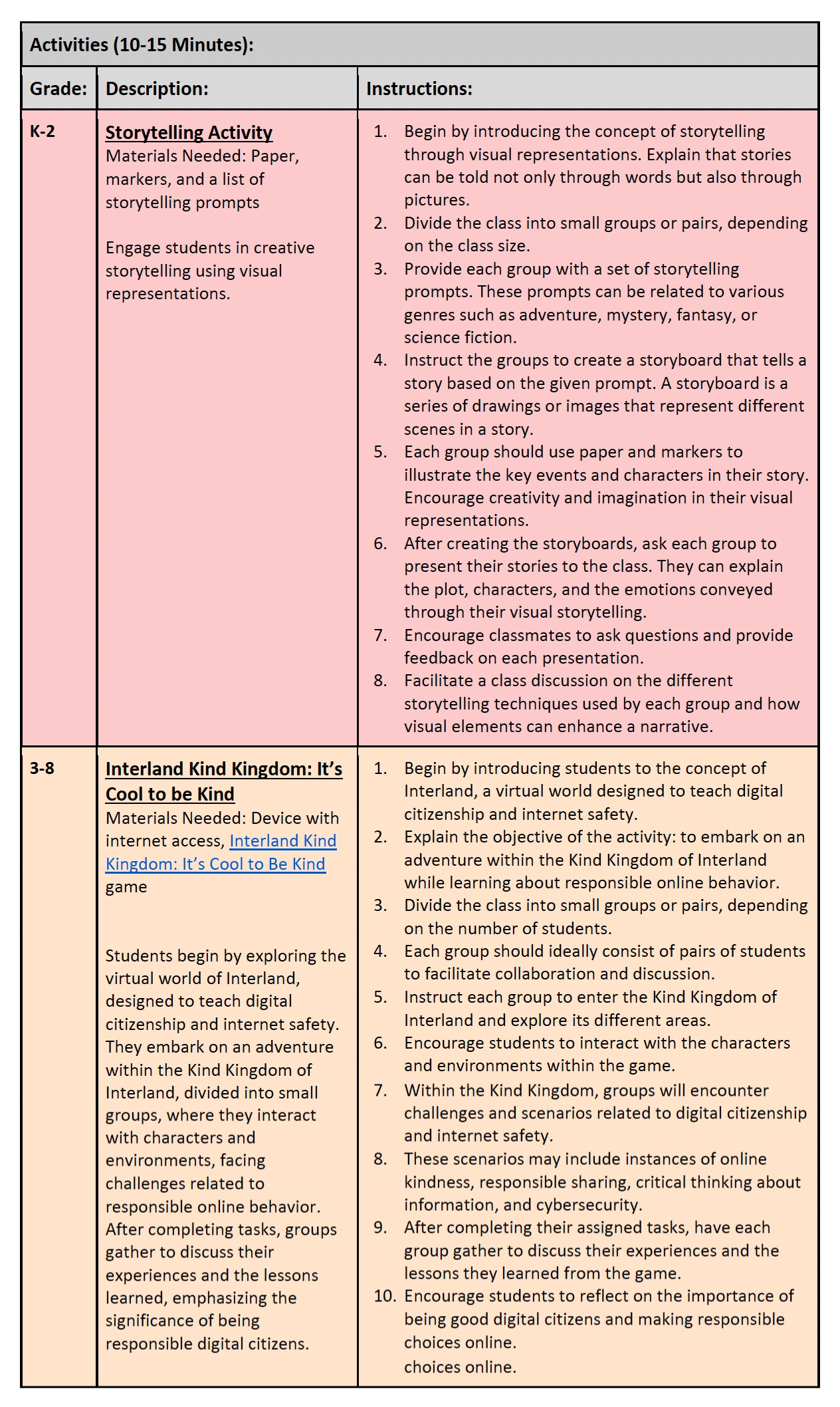}
    \caption{(Cont.) Final lesson plan design for lesson 1 (day 1) in module 1: Digital Citizenship.}
    \label{fig:finallesson2}
\end{figure}


\end{document}